\documentclass[11pt]{article}

\usepackage{soul}
\usepackage{wrapfig}
\usepackage{jcappub} 
\usepackage{graphicx}
\usepackage{lipsum}
\usepackage{tabularx}
\usepackage{textcomp, gensymb}
\usepackage{multirow}
\usepackage{float}
\usepackage{mathtools}
\usepackage{comment,cancel}
\usepackage{indentfirst}
\usepackage{enumerate}
\usepackage[T1]{fontenc}
\usepackage{booktabs}
\usepackage[dvipsnames,table]{xcolor}
\usepackage{hyperref}
\newcommand{\appref}{%
  {\hypersetup{linkcolor=black}\hyperref[app:A]{Appendix}}%
}
\usepackage{xfrac}
\usepackage{placeins}

\newcommand{\ryotext}[1]{\emph{\color{purple} #1}} 
\title{The contribution from small scales on two-point shear analysis: comparison between power spectrum and correlation function}

\author[a,b]{Jo\~ao Ferri}
\author[b]{, Elisa G. M. Ferreira}
\author[c, b]{, Ryo Terasawa}

\affiliation[a]{\small Departamento de F\'{\i}sica Matem\'atica, Instituto de F\'{\i}sica, Universidade de S\~ao Paulo,\\ R.  do  Mat\~ao  1371,  05508-090,  S\~ao Paulo, SP, Brazil}

\affiliation[b]{\small Kavli Institute for the Physics and Mathematics of the Universe (WPI), 
The University of Tokyo Institutes for Advanced Study (UTIAS), The University of Tokyo, Chiba 277-8583, Japan}

\affiliation[c]{Center for Frontier Science, Chiba University, 1-33 Yayoi-cho, Inage-ku, Chiba 263-8522, Japan}

\emailAdd{joao.vitor.ferri@usp.br}
\emailAdd{elisa.ferreira@ipmu.jp}
\emailAdd{teraryo1317@gmail.com}

\abstract{A known problem in cosmic shear two-point statistics is the apparent inconsistency between analyses performed in harmonic space (power spectrum) and real space (angular correlation). This arises mainly from two factors: first, scale cuts in one space correspond to soft cuts in the other, as the relationship between the two spaces is mediated by Bessel functions. For the same reason, astrophysical effects that are compact in one space may not be in the other, which can lead to biased parameter estimates. In this paper, we argue that these two statistics are complementary: we expect a robust theory to provide consistent constraints regardless of the chosen scale cuts. We present the consequences of pushing our analysis to smaller scales in both spaces, accounting for different models of Intrinsic Alignment and Baryonic Feedback in HSC Y3 data: we find that the harmonic-space analysis is significantly less sensitive to the specific modeling of small-scale physics, with model-choice-driven biases in $S_8$ being 2-3 times smaller than in real space. We show that using a flexible, simulation-based emulator for baryonic feedback (\texttt{BACCO}) in combination with the TATT model for intrinsic alignments provides the most consistent cosmological constraints between the two spaces when pushing to the smallest scales. In contrast, the standard \texttt{HMCode-2016} model results in a $\sim 1.1\sigma$ tension between the two statistics. While harmonic space appears more robust for cosmological inference given current model uncertainties, real-space analyses offer a clearer separation of baryonic effects and will play a crucial role in distinguishing between baryonic feedback models in upcoming surveys.

}


\makeatletter
\gdef\@fpheader{}
\makeatother

\begin{document}

\maketitle
\flushbottom

\section{Introduction}
\label{sec:introduction}

\noindent Weak galaxy lensing is a powerful, unbiased probe of the large-scale structure of the Universe (see \cite{bartelmann2016weakgravitationallensing} for a review). It consists of mapping the distortions (or \textit{shear}) imprinted by gravitational lensing on the observed shapes of galaxies. A particularly important summary statistic coming from it is $S_8 \equiv \sigma_8 \left( \frac{\Omega_m}{0.3} \right)^{0.5}\,,$ which captures the main degeneracy direction constrained by cosmic shear surveys. Demonstrating consistency of $S_8$ measurements across different probes is therefore a key test of the standard cosmological model.

Most of the constraining power on $S_8$ can be extracted from the two-point statistics of the shear field, either in real space (leading to the two-point correlation functions $\xi^{ij}_\pm(\theta)$) or in harmonic space (leading to the E-mode angular power spectrum $C_\ell^{ij}$). Here, the indices $i,j$ refer to pairs of redshift bins.

In practice, when comparing observed shear with theoretical predictions, it is necessary to impose \textit{hard} scale cuts. At large scales, cuts are driven by observational limitations such as the survey footprint. At small scales, they arise from theoretical modeling uncertainties related to nonlinear physics, baryonic effects \cite{Schneider_2020,Chisari_2019}, and other systematics, such as intrinsic alignments \cite{Lamman_2024,Kiessling_2015,Kirk_2015,Joachimi_2015}. The choice of scale cuts also depends on the domain of analysis: harmonic space (where cuts are applied in multipole $\ell$) or real space (where cuts are applied in angular separation $\theta$).  

Stage-III surveys such as HSC \cite{Hikage_2019,Hamana_2020} and KiDS \cite{Hildebrandt_2016,K_hlinger_2017} initially reported inconsistencies between real- and harmonic-space cosmic shear analyses, which in turn biased their $S_8$ constraints. Several strategies to reconcile these statistics have been discussed in the literature. For instance, Doux et al.~\cite{Doux_2021} adopted scale cuts in DES that were consistently defined in Fourier space through a common $k_{\rm max}$, which yielded consistent results between the two domains. Alternatively, Park et al.~\cite{park2025matchingcosmicshearanalysis} showed that uncertainties in real-space scales can be propagated into harmonic space (and vice versa), which effectively suppresses the counterpart measurement and removes relative biases. As already emphasized by Doux et al, this trade-off is unavoidable: hard cuts in $\theta$ correspond to soft cuts in $\ell$, and vice versa.  

Moreover, different physical effects manifest differently in the two domains. For example, Terasawa et al.~\cite{terasawa2025equivalencegalaxyangularcorrelation} highlighted that the impact of primordial non-Gaussianity (PNG) is compact in harmonic space, producing a characteristic signature at low multipoles. In real space, however, the same effect spreads across both large and small angular scales, creating degeneracies with other parameters such as $S_8$. This illustrates that one domain can be more sensitive to a given effect than the other, depending on the scale cuts applied. Consequently, analyzing data in both spaces provides complementary information and is crucial for mitigating degeneracies and biases.  

In recent years, galaxy weak lensing surveys have generally reported systematically lower values of $S_8$ compared to the Cosmic Microwave Background (CMB). 
\ryotext{Surveys such as DES-Y3 \cite{blue_shear}, KiDS-Legacy \cite{kids-legacy} and re-analysis of HSC-Y3 \cite{CDJ_2025b} have alleviated this so-called $S_8$ tension with Planck 2018, each through different methodological choices. 
}

Terasawa et al.~\cite{terasawa2024exploringbaryoniceffectsignature} (T24 hereafter) showed that dark matter-only model extending the two-point correlation function (2PCF) analysis to the smallest available angular scales in HSC-Y3 (down to $\theta = 0.28$ arcmin) yields an $S_8$ constraint consistent with the official results~\cite{li2023hsc}.
As an analysis that neglects baryonic feedback is expected to bias the inferred $S_8$ low, one should expect that there is no strong sign of feedback in HSC data, and/or it is not relevant given our current precision. Garcia-Garcia et al.\cite{Garc_a_Garc_a_2024} (GG24, hereafter) analyses with the HSC-Y1 power spectrum seems to say the opposite: there is very strong feedback in HSC-Y1 data, according to their model for baryonic suppression.  

It should be noted that the analysis of T24~was carried out exclusively in real space, leaving open the question of its robustness when translated to harmonic space. The goal of this work is then to explore the relation between real- and harmonic-space analyses when extending the scale cuts, and to assess the impact of incorporating more flexible baryonic feedback models in the HSC Year 3 weak lensing data.  

The remainder of this paper is structured as follows. In Section~\ref{sec:theory}, we briefly review the mathematical framework connecting the observed shear to the two-point statistics $C_\ell$ and $\xi_\pm$. Section~\ref{sec:data} describes the HSC-Y3 data. Section~\ref{sec:methods} presents our analysis methodology, including the pipeline, scale cuts, and modeling choices. Results are discussed in Section~\ref{sec:results}, and we conclude in Section~\ref{sec:conclusions}.

\section{Theory}
\label{sec:theory}
\subsection{Weak Lensing Shear and Two-Point Statistics}
\noindent Weak gravitational lensing refers to the subtle, coherent distortions of the observed shapes of distant galaxies due to the deflection of light by the matter field it passes through. These distortions can be quantified in terms of the \textit{shear field}, a spin-2 field describing the anisotropic stretching of galaxy images. This stretching can be comprised by a \textit{distortion matrix}:

\begin{equation}
\mathbf{A} = 
\begin{pmatrix}
1 - \kappa - \gamma_1 & -\gamma_2 \\
-\gamma_2 & 1 - \kappa + \gamma_1
\end{pmatrix},
\end{equation}
where $\kappa$ is the \textit{convergence}, responsible for isotropic magnification, and $\gamma_1$, $\gamma_2$ are the two components of the shear field, responsible for anisotropic stretching. In Harmonic space, the relation between convergence and shear is given by:

\begin{align}
\tilde{\gamma}_{1\,\ell} &= \tilde{\kappa}_{\ell} \cos 2\varphi_{\ell}, \\
\tilde{\gamma}_{2\,\ell} &= \tilde{\kappa}_{\ell} \sin 2\varphi_{\ell},
\end{align}
where $\varphi_{\ell}$ is the polar angle of the 2D wave vector $\boldsymbol{\ell}$.

Given that individual galaxy shapes are intrinsically noisy due to their unknown intrinsic ellipticities, the cosmological information contained in weak lensing is typically extracted statistically via two-point correlation functions of the shear field. These two-point functions capture most of the cosmological signal, since on large, quasi-linear scales the shear field can be assumed to be a Gaussian random field, for which two-point statistics fully characterize the field.

In real (angular) space, the two-point shear correlation functions are defined as:

\begin{equation}
\xi_{\pm}(\theta) = \left\langle \gamma_t(\boldsymbol{\theta}') \gamma_t(\boldsymbol{\theta}'+\boldsymbol{\theta}) \right\rangle 
\pm \left\langle \gamma_\times(\boldsymbol{\theta}') \gamma_\times(\boldsymbol{\theta}'+\boldsymbol{\theta}) \right\rangle,
\end{equation}
where $\gamma_t$ and $\gamma_\times$ are the tangential and cross components of the shear relative to the separation vector between pairs of galaxies.

In Harmonic space, the shear field is decomposed into $E$-mode and $B$-mode components, analogous to electric and magnetic fields. The $E$-mode arises from the scalar gravitational potential, while the $B$-mode is expected to vanish in the absence of systematics and beyond higher-order lensing effects. The power spectrum of the $E$-mode component is defined as:

\begin{equation}
\left\langle \tilde{\gamma}^E_{\ell} \, \tilde{\gamma}^{E*}_{\ell'} \right\rangle = (2\pi)^2 \delta_D({\ell} - {\ell}') C_{\ell}^{EE},
\end{equation}
where $\tilde{\gamma}^E_{\ell} $, the gradient-like component of the shear, is exactly the convergence field $\tilde{\kappa}_{\ell}$.

The connection between the real-space and harmonic-space descriptions of the shear field mirrors that of the standard two-point analysis of isotropic random fields, where the correlation function $\xi(r)$ and the power spectrum $P(k)$ form a Fourier pair (with Bessel functions in the 2D case):

\begin{align}
\label{eq:xi_plus}
\xi_+(\theta) &= \int_0^\infty \frac{\ell \, d\ell}{2\pi} \, C_{\ell}^{EE} \, J_0(\ell \theta), \\
\label{eq:xi_minus}
\xi_-(\theta) &= \int_0^\infty \frac{\ell \, d\ell}{2\pi} \, C_{\ell}^{EE} \, J_4(\ell \theta).
\end{align}

In practice, the two-point correlation functions $\xi_\pm(\theta)$ are not computed directly from the continuous shear field, but estimated from discrete galaxy samples using pair-count estimators. These estimators inherently impose scale cuts -- set by the finite survey geometry, masking, and angular binning -- which define the smallest and largest separations that can be reliably measured. As a result, the observed $\xi_\pm(\theta)$ contain built-in limitations absent from the theoretical, continuous power spectrum $C_\ell^{EE}$, which is defined over all multipoles. 

Moreover, the relations above highlight the complementarity of the two approaches: while $\xi_{\pm}(\theta)$ directly measure shear correlations as a function of angular separation, the power spectrum $C_{\ell}^{EE}$ quantifies the distribution of power across angular frequency modes. The oscillatory nature of the Bessel functions implies that a sharp scale cut in one space results in a broad, “soft” cut in the other, a fact that becomes particularly important when modeling small-scale effects such as baryonic feedback and intrinsic alignments.

\subsection{Intrinsic Alignment}
\noindent One of the main astrophysical systematics in weak lensing analyses is the phenomenon of \textit{Intrinsic Alignment} (IA), where the intrinsic shapes of galaxies are not randomly oriented but instead respond coherently to the surrounding large-scale structure (see Refs. \cite{Lamman_2024,Kiessling_2015,Kirk_2015,Joachimi_2015} for a review). These correlations can bias cosmic shear measurements if not properly modeled and marginalized over.

Intrinsic Alignments manifest primarily through two types of correlations: \textit{Intrinsic-Intrinsic} (II) and \textit{Gravitational-Intrinsic} (GI) terms. II accounts for the alignment between the intrinsic ellipticities of nearby galaxies, whereas GI is the correlation between the intrinsic ellipticity of a galaxy and the gravitational shear experienced by a background galaxy. 

Here, we expose two widely used models in modern cosmic shear analyses: \textit{Non-Linear Alignment} (NLA) and \textit{Tidal Alignment and Tidal Torquing }(TATT) models.

\subsubsection*{NLA Model}
\noindent The NLA model is an extension of the linear alignment model \cite{2004PhRvD..70f3526H}, incorporating the non-linear matter power spectrum to better capture alignments on small scales. The intrinsic alignment power spectra are modeled as:

\begin{align}
P_{\mathrm{II}}(k, z) &= C_1^2(z) \, P_{\delta}(k, z), \\
P_{\mathrm{GI}}(k, z) &= C_1(z) \, P_{\delta}(k, z),
\end{align}
where $P_{\delta}(k, z)$ is the non-linear matter power spectrum, and 

\begin{align}
\label{eq:TA_term}
    C_1(z) = -A_{1} \frac{\bar{C_1}\rho_{\mathrm{crit}} \, \Omega_m}{D(z)} \left( \frac{1+z}{1+z_{piv}} \right)^{\eta_1}\,.
\end{align}

The two parameters of this theory are $A_{1}$, the intrinsic alignment amplitude parameter, and $\eta_{1}$, which allows for redshift evolution of the alignment strength. $\bar{C_1}$ is a normalization constant set to $5 \times 10^{-14} \, h^{-2} M_{\odot}^{-1} \mathrm{Mpc}^3$, and $z_{piv}$ is a pivotal redshift usually set to $0.62$ in DES-and HSC-like experiments.

These IA power spectra enter both real-space and harmonic-space two-point functions. In harmonic space, for example, the total observed shear power spectrum becomes:

\begin{equation}
C_{\ell}^{\mathrm{obs}} = C_{\ell}^{\mathrm{GG}} + C_{\ell}^{\mathrm{II}} + C_{\ell}^{\mathrm{GI}},
\end{equation}
and in real space, the IA terms contribute similarly to the shear two-point correlation functions \(\xi_{\pm}(\theta)\) via the Bessel transforms \eqref{eq:xi_plus}-\eqref{eq:xi_minus}.

\subsubsection*{TATT Model}
\noindent The TATT model \cite{Blazek_2019} generalizes IA modeling by including both tidal alignment and tidal torquing mechanisms, aiming to capture alignments for both elliptical and spiral galaxies. The intrinsic shape field \(\gamma^{I}\) is modeled as a perturbative expansion in the tidal field, up to quadratic order:

\begin{align}
\gamma^{I}_{ij} &= C_1(z)\, s_{ij} + C_2(z)\, \sum_ks_{ik}s_{kj} +b_{TA}\,C_1(z)\,\delta s_{ij} + \cdots,
\end{align}
where \(s_{ij}\) is the tidal shear tensor, and \(C_1, C_2\) are alignment coefficients controlling the amplitude of the linear (alignment) and quadratic (torquing) terms, respectively. We also allow for a bias $b_{TA}$ in the alignment term. While $C_1(z)$ has the same expression of \eqref{eq:TA_term}, $C_2(z)$ is given by

\begin{align}
    \label{eq:TT_term}
    C_2(z)=5\,A_2\frac{\bar{C_1}\rho_\mathrm{crit}\,\Omega_M}{D^2(z)}\left(\frac{1+z}{1+z_{piv}}\right)^{\eta_2}\,.
\end{align}

Essentially, the TATT model has up to 5 parameters: $A_1,A_2,\eta_1,\eta_2,b_{TA}$. While the NLA model is simpler and widely used, the TATT model offers greater flexibility, particularly in joint analyses involving both early-type (elliptical) and late-type (spiral) galaxies.

Recent results from the DES Y3 ``Blue Shear'' analysis \cite{blue_shear} provided an instructive example of how IA can be an important contaminant in shear analyses. By constructing a high-purity sample of blue galaxies (which are empirically found to exhibit negligible intrinsic alignments), the DES team performed a cosmic shear analysis largely insensitive to IA modeling assumptions. They showed that cosmological constraints derived from the blue sample are both more stable across IA model choices (varying by only $\sim 0.5\sigma$ in $S_8$) and yield improved agreement with Planck CMB measurements compared to the full or red samples. The resulting constraints, $S_8 = 0.822^{+0.019}_{-0.020}$ and $\Omega_m = 0.268^{+0.031}_{-0.056}$, are consistent with Planck within $1\sigma$, and the Bayesian evidence strongly favors the absence of IA in this sample.

\subsection{Baryonic Physics Correction}

\noindent Weak gravitational lensing probes the \textit{total} matter distribution in the Universe, including both dark matter and baryonic matter. While the large-scale distribution of dark matter is well described by gravitational physics alone, baryonic processes -- such as gas cooling, star formation, and feedback from supernovae and active galactic nuclei (AGN) -- significantly modify the matter distribution on small, non-linear scales. High-resolution hydrodynamical simulations such as the OWLS \cite{2010MNRAS.402.1536S,Le_Brun_2014}, Illustris \cite{stu1654,Vogelsberger_2014}, and EAGLE \cite{2015MNRAS.446..521S} projects revealed that these processes redistribute mass within halos, alter halo density profiles, and suppress or enhance the matter power spectrum at scales of $k \gtrsim 0.5$-$1\, h\, \mathrm{Mpc}^{-1}$. Neglecting these effects in cosmological parameter inference, particularly when including small-scale data, can lead to biased estimates of key parameters such as $S_8$ and $\sigma_8$ \cite{Semboloni_2011,Semboloni_2013,Zentner_2013}. 

With the precision achieved by current surveys such as HSC and the upcoming Stage-IV experiments (e.g., Rubin, Euclid, and Roman), the statistical uncertainties are now small enough that baryonic effects become a dominant source of modeling error on the small scales probed by these datasets. While one could in principle remove these scales through conservative scale cuts, doing so discards valuable cosmological information. As the quality and depth of lensing data continue to improve, properly modeling baryonic feedback is therefore essential to fully exploit the constraining power of weak-lensing measurements and to avoid systematic biases in cosmological inference.

To incorporate these effects in shear analysis, some developed strategies are:

\subsubsection*{Model predictions}
\noindent While hydrodynamical simulations offer the most direct means of assessing the impact of baryonic physics on the matter power spectrum, their computational cost makes them impractical for incorporation into cosmological parameter inference. As a result, a range of analytic and semi-analytic models have been developed to approximate baryonic effects in a computationally efficient and physically interpretable way.

Notable examples in the literature include Zentner et al. (2013) \cite{Zentner_2013}, Schneider \& Teyssier (2015) \cite{Schneider_2015}, and Mead et al. (2016) \cite{Mead_2015}. In particular, Mead et al. 2016's \texttt{HMCode} \cite{Mead_2015,Mead_2016} has been adopted in several Stage-III survey analyses, including KiDS-450 \cite{Hildebrandt_2016}, DES Y1 \cite{Troxel_2018}, and HSC Y3 \cite{li2023hypersuprimecamyear3,Dalal_2023}.

To account for baryonic suppression, \texttt{HMCode} modifies two aspects of the halo model:

\begin{itemize}
    \item The \textit{halo concentration-mass relation}, through a parameter $A_{\mathrm{baryon}}$:
\end{itemize}

\begin{equation}
    \label{eq:A_baryon}
    c(M,z)\to\frac{A_{\mathrm{baryon}}}{3.13}\,c_{DMO}(M,z)
\end{equation}

\begin{itemize}
    \item The \textit{halo "bloating"} parameter, denoted as $\eta_b$, which controls the scale-dependence of the halo density profiles:
\end{itemize}

\begin{equation}
    \label{eq:eta_b}
    u(k|M,z) \rightarrow u(k/\eta_b|M,z).
\end{equation}

Following Joudaki et al. \cite{Joudaki_2017}, HSC Y3 adopts KiDS-450 relation between $A_{\mathrm{baryon}}$ and $\eta_b$:
\begin{equation}
    \eta_b = 0.98 - 0.12\,A_{\mathrm{baryon}},
\end{equation}
such that larger values of $A_{\mathrm{baryon}}$ correspond to weaker baryonic suppression. Specifically, the value $A_{\mathrm{baryon}} = 3.13$ recovers the dark-matter-only (DMO) prediction of the matter power spectrum, equivalent to the baseline model without baryonic feedback.  Lower values of $\eta_b$ (associated with stronger baryonic feedback) broaden the halo profiles, which leads to a reduction of small-scale clustering power. $A_{\mathrm{baryon}}$ is treated as a nuisance parameter, typically marginalized over with a conservative prior, while $\eta_b$ is computed from it using the relation above.

\subsubsection*{Simulation-based emulators}

\noindent An alternative to (semi-)analytical models for incorporating baryonic effects in weak-lensing analyses is the use of simulation-based emulators, as they enable reliable predictions down to significantly smaller, non-linear scales. These tools interpolate between outputs of suites of high-resolution hydrodynamical or $N$-body simulations, whose sampling points in parameter space are distributed using space-filling designs such as Latin Hypercube or Sobol sequences, ensuring an efficient and uniform coverage of the cosmological and astrophysical parameter volume. Once trained, the emulator can then accurately predict (i.e., emulate) quantities such as the non-linear matter power spectrum or correlation functions at new parameter combinations, providing a computationally cheap yet precise alternative to running full simulations for each model evaluation. Some examples include \texttt{EuclidEmulator} \cite{euclidemulator2019}, \texttt{CosmicEmu} \cite{Heitmann_2013}, and most recently \texttt{BACCO} \cite{Angulo_2021}, which specifically incorporates baryonic feedback effects using the \textit{baryonification} approach.

\subsubsection*{Baryonification and the \texttt{BACCO} Emulator}
\noindent The baryonification technique \cite{Schneider_2019} provides an efficient way to approximate baryonic effects by post-processing the output of dark-matter-only $N$-body simulations. It modifies the spatial distribution of matter by shifting particles and adjusting halo density profiles according to physically motivated prescriptions calibrated on hydrodynamical simulations. In practice, baryonification mimics the redistribution of matter driven by feedback processes such as AGN and stellar winds, broadens or reshapes halo profiles to account for baryonic physics, and, when necessary, adds a diffuse gas component at large radii. This approach captures the main effects of baryons on the matter power spectrum at a fraction of the computational cost of full hydrodynamical simulations.

Building on this framework, the \texttt{BACCO} emulator \cite{Angulo_2021} combines a suite of high-accuracy $N$-body simulations with baryonification to reproduce the non-linear matter power spectrum across a wide range of cosmological and feedback scenarios. Rather than simply interpolating between discrete models, \texttt{BACCO} is trained to emulate the impact of different cosmologies and baryonic feedback strengths with high precision and computational efficiency.

Within the \texttt{BACCO} framework, baryonic effects are governed by a small set of physically motivated parameters. The parameter $M_c$ sets the halo mass above which feedback significantly alters density profiles, while $\eta$ controls how far gas is ejected from halos. The slope of the mass dependence is set by $\beta$, and $M_{1,z_0,{\rm cen}}$ determines the mass scale associated with the central galaxy component. The parameters $\theta_{\mathrm{inn}}$ and $\theta_{\mathrm{out}}$ define the inner and outer radial ranges where baryonic modifications are applied, and $M_{\mathrm{inn}}$ regulates the mass redistribution in the inner halo region. Together, these parameters allow \texttt{BACCO} to reproduce a broad spectrum of baryonic scenarios, ranging from strong AGN feedback to nearly dark-matter-only behavior.

\section{Scale Cuts}
\label{subsec:scale_cuts}
\noindent A more extreme strategy, often used along with forward modeling baryonic effects and intrinsic alignment, is to cut altogether the data at scales $k \gtrsim 1\, h\, \mathrm{Mpc}^{-1}$, where these effects are stronger. However, several problems arise with scale cuts.

\begin{figure}[ht!]
    \centering
    \hspace{-1cm} 
    \includegraphics[width=1.0\linewidth]{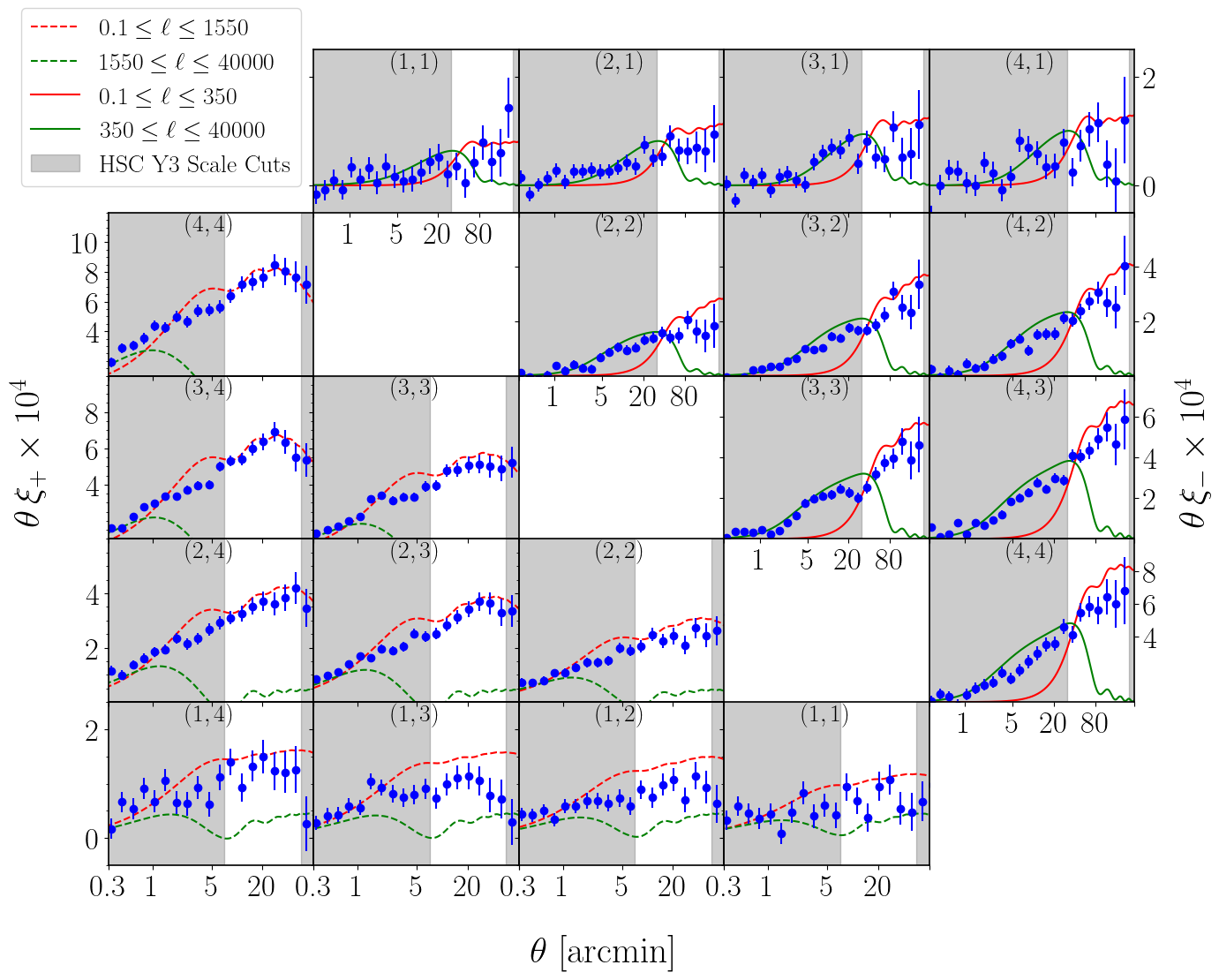}
    \caption{The full $\xi_\pm$ data vector (in blue) for the HSC Y3 measurements. The red and green lines are the convolutions of $C_\ell$ to $\xi$ via Eqs. \eqref{eq:xi_plus}-\eqref{eq:xi_minus}, computed with the best-fit values from \cite{li2023hypersuprimecamyear3} and different multipole intervals. The shaded areas correspond to scales cut away from the official HSC-Y3 analysis.}
    \label{fig:xi_pm}
\end{figure}

Figure \ref{fig:xi_pm} shows that approximately \texttt{2/3} of all data points from the official HSC-Y3 analysis are thrown away ($\theta<7\,\,\rm arcmin$ for $\xi_-$ and $\theta<31\,\,\rm arcmin$ for $\xi_+$). While these points may be prone to systematics, they contain valuable cosmological data that can improve our constraints.

\begin{figure}
    \centering
    \includegraphics[width=1\linewidth]{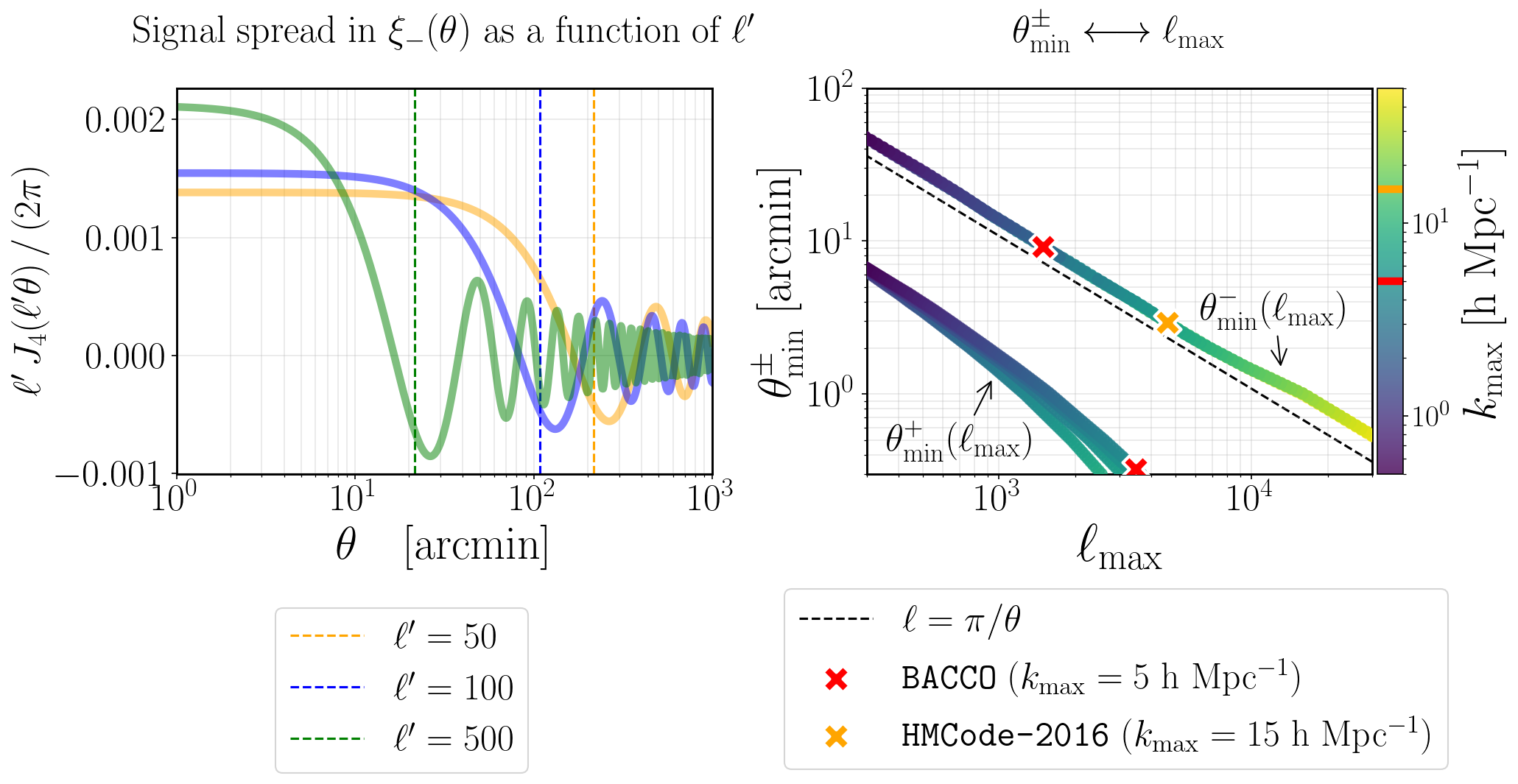}
    \caption{(Left) Compact signals in harmonic space (given by $\delta_D(\ell-\ell^\prime)$) as seen in real space. (Right) Interpolation of Doux et al. \cite{Doux_2021} $k_{\rm max}\longleftrightarrow(\theta^\pm_{\rm min},\ell_{\rm max})$. The upper scatter corresponds to $\theta^-_{\rm min}\longleftrightarrow\ell_{\rm max}$, and the lower scatter corresponds to $\theta^+_{\rm min}\longleftrightarrow\ell_{\rm max}$. For visualization purposes, we also plot the usual Nyquist frequency $\ell=\pi\,/\,\theta$ .
    }
    \label{fig:doux}
\end{figure}

A second problem comes from the fact that there is no direct correspondence between a scale cut in one space to a scale cut in the other. For instance, let us take Eqs. \eqref{eq:xi_plus}-\eqref{eq:xi_minus}. The left plot on Figure \ref{fig:doux} shows what happens when $C_\ell$ is a compact signal at some $\ell^\prime$ (given by a Dirac delta): its sharp signal in harmonic-space gets diluted in real space. In special, a great portion of the signal shifts to \textit{bigger} scales (bigger $\theta$'s in comparison with the usual relation $\theta=\pi/\ell^\prime$). This is specially evident on the upper triangle of Fig. \ref{fig:xi_pm}, where multipoles bigger than $350$ (roughly $\theta=31\,\,\rm arcmin$) still contribute significantly to $\xi_-$ at higher scales. At the same time, a non-negligible portion of the information also gets distributed to \textit{smaller} scales. We refer to Figure 14 from GG24\cite{Garc_a_Garc_a_2024}, where it is evident that a full characterization of $\xi_-(\theta)$ at $1\%$ error requires an integration up to $\ell_{\rm max}$ much higher than the naive relation $\pi/\theta$; it is needed, in fact, 10 times that number of multipoles. In a current scenario, however, we do not need to integrate up to such a high multipole: as the errors of $\xi_{\pm}$ range from $50\%$ to $20\%$ in measurements below $\theta=7\rm\,\,arcmin$, information from multipoles lower than the naive relation $\pi/\theta$ are smeared out. We will review this statement again in Section \ref{sec:results}, testing it with the data.

From the data perspective, this also reflects at the time of making a correspondence between data cuts in one space and another. For instance, we consider Doux et al. \cite{Doux_2021} method to build a $\theta-\ell$ correspondence: it starts from a physical mode cut-off $k_{\rm max}$, and then we impose that

\begin{equation}
\label{eq:Doux}
    \int^{ln\,k_{\rm max}}_{-\infty}\bigg|\frac{\partial\,ln\,\rm X_{\theta,\ell}}{\partial\,ln\,k}\bigg|\,d\,ln\,k<0.95\,\,,
\end{equation}
where $X_{\theta,\ell}$ may be either $\xi_{\pm}(\theta_{\rm min})$ or $C_{\ell_{\rm max}}$ -- expressed in terms of $k$ by the usual relation $k=(\ell(\theta)+1/2)\,/\,\chi(z)$\,. Equation \ref{eq:Doux} means that the contribution of physical modes of scales $k>k_{\rm max}$ only contribute to $5\%$ of the \textit{total} power of $\xi_\pm(\theta_{\rm min})/ C_{\ell_{\rm max}}$. The right panel of Figure $\ref{fig:doux}$ shows an interpolation of the $k_{\rm max}\times\theta_{\rm min}$ and $k_{\rm max}\times\ell_{\rm max}$ relations, translated into a $\theta_{\rm min}$ versus $\ell_{\rm max}$ relationship (given a typical $\Lambda$CDM cosmology and HSC-like redshift bin distributions). The plot highlights the inherent differences between the two-point correlation functions: for $\xi_{-}$, the relationship is roughly $\theta^{-}_{\rm min}\simeq1.2\,\pi\,/\,\ell_{\rm max}$, whereas for $\xi_{+}$ the relation is much lower, $\theta^{+}_{\rm min}\simeq0.3\,\pi\,/\,\ell_{\rm max}$. This shows that $\xi_{-}$ is roughly four times more sensitive to small-scale physics (higher $k_{\rm max}$) than $\xi_{+}$ (as previously pointed out by T24), which directly dictates the chosen scale cuts. The markers in this panel illustrate the maximum applicable $k$ range for the baryonic feedback models: $\texttt{HMCode-2016}$ and $\texttt{BACCO}$ models were calibrated up to $k_{\rm max}$ values of $15\,h\,\text{Mpc}^{-1}$ and $5\,h\,\text{Mpc}^{-1}$, respectively. It is important to note that although the nonlinear dark-matter-only (DMO) matter power spectrum in \texttt{HMCode-2016} is calibrated up to $15\,h\,{\rm Mpc}^{-1}$, its baryonic modeling is only reliable up to $k_{\rm max} \sim 5\,h\,{\rm Mpc}^{-1}$ due to the absence of the characteristic upturn at high $k$ (see Fig. 5 of Mead et al.~\cite{Mead_2021}). 

In light of this, we can safely apply both $\texttt{BACCO}$ and $\texttt{HMCode-2016}$ across the full range of $\xi_{+}$ scales explored. However, the same cannot be stated for $\xi_{-}$ analyses below $\theta_{\min}\sim 3$ arcmin or $C_{\ell}$ analyses with $\ell$ higher than $\sim 5000$. For $k \gtrsim 5\,h\,{\rm Mpc}^{-1}$, we effectively use extrapolated model that mimics a WDM-like suppression, rather than a physically motivated baryonic effect.

\begin{figure}
    \centering
    \includegraphics[width=1\linewidth]{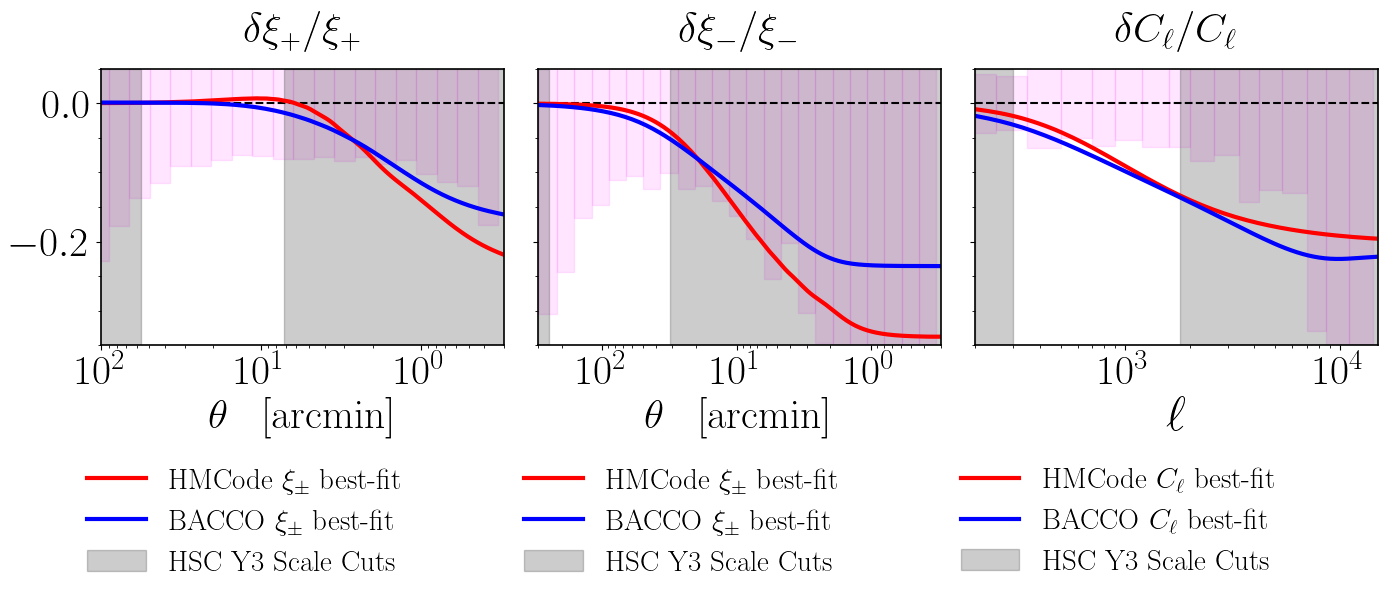}
    \caption{Impact of baryonic feedback on the dark-matter-only correlation functions. We show the fractional suppression predicted by the \texttt{HMCode-2016} and \texttt{BACCO} models, each fitted to HSC-Y3 data. The magenta region marks the sensitivity range of the HSC-Y3 measurements.}
    \label{fig:suppression}
\end{figure}

Finally, in Figure~\ref{fig:suppression} we explicitly illustrate the impact of baryonic feedback by showing the fractional suppression relative to the dark-matter–only predictions, defined as $(X - X_{\mathrm{DM}})/X_{\mathrm{DM}} \equiv \delta X / X$, for the auto-correlations of the last redshift bin. Although this bin is less sensitive to nonlinearities than the lower-redshift bins, it provides the highest SNR due to the nature of the shear signal. Here, $X$ corresponds to either $\xi_\pm$ or $C_\ell$. By fitting the \texttt{HMCode-2016} and \texttt{BACCO} models to the \textit{full} HSC-Y3 data, we find a fundamental difference between the real-space and harmonic-space representations: since baryonic feedback acts as a localized effect in real space, the inverse Hankel transforms in Eqs.~(\ref{eq:xi_plus})-(\ref{eq:xi_minus}) spread this suppression over a broader range of multipoles. This is evident in the $C_\ell$ panel, where even within the official scale cuts we observe a significant suppression that exceeds the survey sensitivity, whereas the same effect is less pronounced in real space. Although the harmonic space representation may smear certain model-dependent, localized effects, real space analysis simplifies the process of imposing physical scale cuts for isolating these effects.

While the data cuts provided by \eqref{eq:Doux} may currently give consistent results across real and harmonic spaces, the same may not happen as we push our scales to regions in which localized effects (such as baryonic feedback in real space) take place. This is because even when making a \textit{data} cut at a certain $\ell_{\rm max}$ (and a correspondent $\theta_{\rm min}$), the relations \eqref{eq:xi_plus}-\eqref{eq:xi_minus} also imply that $\xi_{\pm}(\theta_{\rm min})$ (particularly $\xi_-$) has much more small-scale information than $C_{\ell_{\rm max}}$. Therefore, we cannot always rely on a data cut matching to access the robustness of our results -- specially in stage-IV surveys, where our measurements will have a higher signal-to-noise. This is where we argue that we may orient the robustness of our analyses not by a data cut matching, but by our theoretical input, which should provide us similar constraints no matter the space in which we are making our analysis. To test this assumption, we are going to test different models to baryonic feedback and intrinsic alignment in HSC-Y3 data. 

\section{Data}
\label{sec:data}
\noindent The HSC Y3\cite{Aihara_2017,Aihara_2022} shear catalog covers an effective area of approximately 416~deg$^2$ and contains about 25 million galaxies, yielding an effective number density of $n_{\mathrm{eff}} \approx 15~\mathrm{arcmin}^{-2}$ within the redshift range $0.3 \leq z < 1.5$ \cite{li2023hypersuprimecamyear3}. The sample is defined by a magnitude cut of $i\leq24.5$ and is divided into four tomographic redshift bins with edges at $z_\mathrm{phot} \in \{0.3, 0.6, 0.9, 1.2, 1.5\}$. The redshift distributions $n(z)$ are calibrated using a combination of COSMOS-2015 \cite{Cosmos2015} and clustering redshifts \cite{Mandelbaum_2017}.

For the analysis in harmonic space, we utilize the angular power spectra $C_\ell$ measurements described in \cite{Dalal_2023}. These spectra are binned into log-spaced bandpowers covering the multipole range $300 \leq \ell \leq 14200$. For the real space analysis, we use the two-point correlation functions, $\xi_{\pm}(\theta)$, measured in T24 \cite{terasawa2024exploringbaryoniceffectsignature} (also see Li et al. (2023) \cite{li2023hypersuprimecamyear3}). The angular binning scheme consists of 21 logarithmically spaced bins between $0.32'$ and $120'$ for $\xi_+$, and 23 bins between $0.32'$ and $216.32'$ for $\xi_-$.

Detailed discussions regarding the data processing, catalog creation, and measurements can be found in Dalal et al. (2023) \cite{Dalal_2023}, T24 \cite{terasawa2024exploringbaryoniceffectsignature}, and Li et al. (2023) \cite{li2023hypersuprimecamyear3}.

\section{Methods}
\label{sec:methods}
\noindent The primary goal of this work is to investigate the impact of including smaller scales in cosmic shear analyses, in both harmonic and real space, and to assess how this affects constraints on cosmological parameters within the $\Lambda$CDM framework, as well as on baryonic feedback and intrinsic alignment (IA) parameters.

Throughout these analyses, the parameter priors are kept consistent with those adopted in the official HSC Year 3 cosmological analysis. We explore a series of scenarios designed to test the interplay between modeling choices and data combinations:

\begin{itemize}
    \item \textbf{HMCode+TATT:} TATT model for intrinsic alignments and \texttt{HMCode-2016} for baryonic feedback (same as official HSC-Y3 analyses);
    \item \textbf{HMcode+NLA:} NLA model for intrinsic alignments and \texttt{HMCode-2016} for baryonic feedback;
    \item \textbf{BACCO+TATT} Replacing \texttt{HMCode-2016} with the \texttt{BACCO} emulator for baryonic effects and nonlinear physics;
    \item \textbf{BACCO+NLA:} NLA model for intrinsic alignments and \texttt{BACCO} for baryonic feedback/nonlinear physics;    
\end{itemize}

For each scenario, we systematically vary the small-scale (lower-limit) cuts on the data while maintaining the same large-scale (upper-limit) cuts as in the official HSC Y3 analysis: $\ell_{\mathrm{min}}=300$ for the harmonic-space analysis, and $\theta_{\mathrm{max}}=56.52'$ for $\xi_{+}$ and $247.75'$ for $\xi_{-}$.

In harmonic space, we explore the following range of values for the upper multipole cut:
\begin{equation}
    \ell_{\mathrm{max}} \in \{2000, 2600, 3400, 4200, 5400, 7000, 8600, 11000, 14200\}.
\end{equation}
Each increment in $\ell_{\mathrm{max}}$ roughly corresponds to the addition of a new $\ell$-bandpower measurement for each auto- and cross-correlation of $C_{\ell}^{EE,ij}$, introducing ten additional data points into the analysis. In real space, we consider the following values for the lower angular scale cut:
\begin{equation}
    \theta_{\mathrm{min}} \in \{0.3', 0.5', 1.0', 1.9', 3.4', 6.2', 8.4'\}.
\end{equation}
These values are applied simultaneously to both $\xi_{+}$ and $\xi_{-}$ measurements.

It is important to note that these scale cuts are fixed across redshift bins, following the convention of the official HSC Y3 analysis. While this is not an optimal cutting strategy -- as the scale at which non-linearities become significant depends on both angular scale and redshift -- it is sufficient for our purposes, where the main objective is to investigate the effect of progressively including smaller scales rather than refining the removal of non-linear contributions.

\subsection{Parameter Inference Pipeline}
\label{subsec:parameter_inference}
\noindent All analyses presented here were performed using the \texttt{CosmoSIS} software framework \cite{Zuntz_2015}, consistent with the official HSC Y3 cosmic shear pipeline. The theoretical model for the linear matter power spectrum was computed with \texttt{CAMB}, using either nonlinear corrections from \texttt{HMCode}'s implementation of \texttt{Halofit} (up to $k=15\,h\,\mathrm{Mpc}^{-1}$), or from \texttt{BACCO} (up to $k=5\,h\,\mathrm{Mpc}^{-1}$). The total power spectrum is then extrapolated to $k=500\,h\,\mathrm{Mpc}^{-1}$ to ensure coverage of the smallest angular scales in our analyses.

We also employ the \texttt{FAST-PT} module \cite{McEwen_2016}, which efficiently computes convolution integrals that arise in perturbation theory -- such as loop corrections and intrinsic alignment contributions -- using fast Fourier transform techniques. This allows us to accurately compute the scale- and redshift-dependent GI and II terms in the matter power spectrum without incurring significant computational overhead.

Following the matter power spectrum calculation:

\begin{enumerate}
    \item We apply photo-$z$ bias shifts derived from the measured redshift distribution in each tomographic bin.
    \item We compute the IA contributions -- II and GI -- to the Fourier-space power spectrum.
    \item These, combined with the lensing GG term and baryonic corrections, are projected into angular power spectra $C_\ell^{EE,\mathrm{obs}}$ including multiplicative shear calibration biases.
\end{enumerate}

For harmonic-space analyses, the theoretical $C_\ell^{EE,\mathrm{obs}}$ vector is directly compared with the public HSC Y3 data in \texttt{cosmosis\_standard\_library}\footnote{https://cosmosis.readthedocs.io/en/latest/}, consisting of 170 bandpower measurements spanning $150 \leq \ell \leq 14200$. In real-space analyses, the projected $C_\ell$ is converted to $\xi_{\pm}(\theta)$ via Eqs. \eqref{eq:xi_plus}-\eqref{eq:xi_minus}, and compared to re-measured data from T24\cite{terasawa2024exploringbaryoniceffectsignature}, which comprises 210 points for $\xi_+$ (from $0.32'$ to $102.10'$) and 230 points for $\xi_-$ (from $0.32'$ to $216.32'$).

We explore the cosmological and nuisance parameter space using the nested sampler \texttt{MultiNest} \cite{Feroz_2009}, which efficiently handles multimodal posteriors and computes Bayesian evidence.  We note that \texttt{MultiNest} can sometimes underestimate posterior uncertainties in high-dimensional problems, and we account for this when interpreting our results.

All priors and fixed parameter values (such as neutrino masses, fixed cosmological constants, and IA/baryonic pivots) match those used in the official HSC Y3 cosmology analysis. The only exception is regarding $A_{baryon}$, where we allow it to span the range $[0.0,3.13]$. Although values $A_{baryon}<2$ may not be physically well-motivated, analyses that present such constraints can be indicative of the need for better modeling. A complete list of sampled parameters, their priors, and fixed parameters is provided in Table \ref{tab:priors}.

\section{Results}
\label{sec:results}
\begin{figure}[h!]
    \hspace{-7em}
    \includegraphics[width=1.2\linewidth]{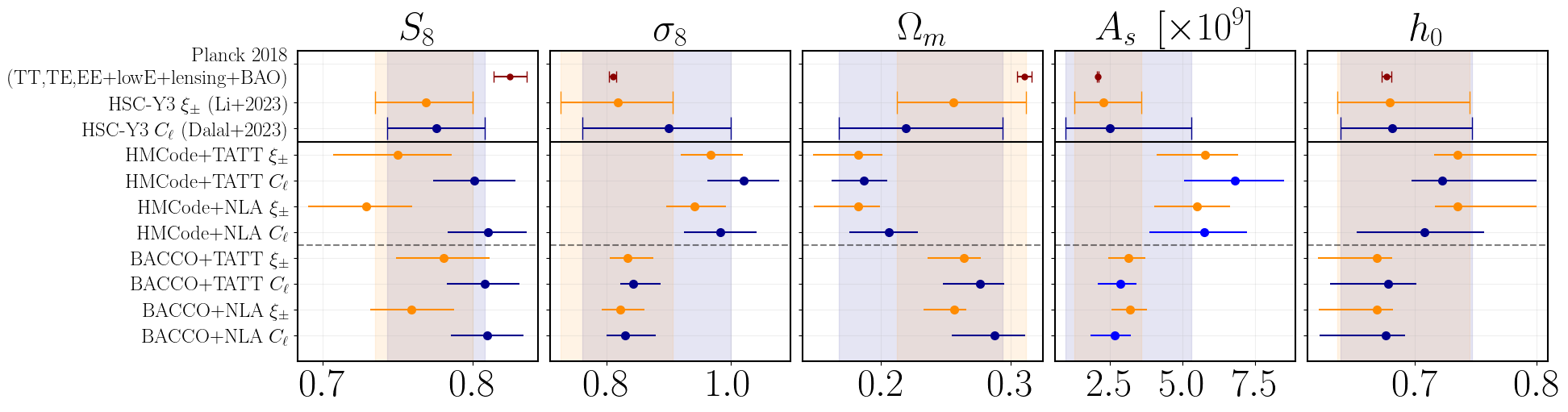}
    \caption{Main cosmological results for all scenarios. For comparison, we show Planck 2018 (in red), and the HSC Y3 harmonic- and real-space analyses in the upper panels. Harmonic-space analyses are shown in blue, whereas the real-space analyses are shown in orange. The middle panels correspond to models employing \texttt{HMCode-2016} for baryonic feedback and nonlinear corrections, while the lower panels use \texttt{BACCO}.
    }
    \label{fig:whisker}
\end{figure}

\noindent We start by reporting our main cosmological results obtained when including all available scales in real and harmonic spaces. Figure \ref{fig:whisker} shows the performance of different models in comparison with the official HSC Y3 and Planck 2018 results. The best-fit values and maximum a posteriori (MAP) estimates are listed in Table \ref{tab:posteriors_summary}. The full 2D posteriors are presented in the \appref. We also note that for all considered scenarios, the posterior means of all parameters differ from their respective MAP values by $\lesssim1\sigma$. 

\begin{table*}[h!] \centering \resizebox{1\textwidth}{!}{ \begin{tabular}{lcccccc} \hline Model & $S_8$ & $\sigma_8$ & $\Omega_m$ & $h_0$ & $A_s\,[\times10^{9}]$\\ \hline \\ HMCode + TATT (Real) & $0.750^{+0.036}_{-0.043}\,(0.736)$ & $0.968^{+0.052}_{-0.049}\,(0.980)$ & $0.182^{+0.019}_{-0.035}\,(0.169)$ & $0.735^{+0.061}_{-0.025}\,(0.735)$ & $5.76^{+1.13}_{-1.66}\,(5.84)$ \\ HMCode + TATT (Harmonic) & $0.801^{+0.027}_{-0.028}\,(0.805)$ & $1.021^{+0.057}_{-0.059}\,(1.085)$ & $0.186^{+0.019}_{-0.025}\,(0.165)$ & $0.722^{+0.063}_{-0.038}\,(0.746)$ & $6.78^{+1.66}_{-1.70}\,(7.80)$ \\ HMCode + NLA (Real) & $0.729^{+0.031}_{-0.038}\,(0.724)$ & $0.941^{+0.051}_{-0.045}\,(0.958)$ & $0.182^{+0.017}_{-0.034}\,(0.171)$ & $0.735^{+0.061}_{-0.025}\,(0.723)$ & $5.49^{+1.13}_{-1.47}\,(6.07)$ \\ HMCode + NLA (Harmonic) & $0.810^{+0.026}_{-0.027}\,(0.810)$ & $0.983^{+0.059}_{-0.058}\,(0.981)$ & $0.206^{+0.023}_{-0.031}\,(0.205)$ & $0.708^{+0.048}_{-0.052}\,(0.732)$ & $5.74^{+1.46}_{-1.86}\,(4.72)$ \\ BACCO + TATT (Real) & $0.781^{+0.031}_{-0.032}\,(0.799)$ & $0.833^{+0.042}_{-0.028}\,(0.851)$ & $0.264^{+0.013}_{-0.028}\,(0.265)$ & $0.669^{+0.022}_{-0.038}\,(0.668)$ & $3.13^{+0.58}_{-0.69}\,(3.33)$ \\ BACCO + TATT (Harmonic) & $0.808^{+0.023}_{-0.025}\,(0.828)$ & $0.843^{+0.043}_{-0.020}\,(0.839)$ & $0.277^{+0.018}_{-0.029}\,(0.292)$ & $0.678^{+0.026}_{-0.043}\,(0.715)$ & $2.87^{+0.55}_{-0.78}\,(2.10)$ \\ BACCO + NLA (Real) & $0.759^{+0.028}_{-0.027}\,(0.783)$ & $0.822^{+0.038}_{-0.030}\,(0.873)$ & $0.257^{+0.009}_{-0.024}\,(0.241)$ & $0.669^{+0.023}_{-0.038}\,(0.739)$ & $3.21^{+0.58}_{-0.65}\,(2.90)$ \\ BACCO + NLA (Harmonic) & $0.810^{+0.024}_{-0.024}\,(0.804)$ & $0.830^{+0.049}_{-0.029}\,(0.891)$ & $0.288^{+0.024}_{-0.033}\,(0.244)$ & $0.676^{+0.023}_{-0.046}\,(0.656)$ & $2.66^{+0.57}_{-0.81}\,(4.13)$ \\\\ \hline \end{tabular} } \caption{Summary of best-fit, 68\% credible intervals, and MAP values for each model.} \label{tab:posteriors_summary} \end{table*}

Our baseline constraint corresponds to the \textbf{BACCO+TATT} scenario, which features the largest number of free parameters:

\begin{equation}
    \begin{gathered}
        S_8\,:\,\,0.808^{+0.026}_{-0.027}\,\,(0.828) \text{ 
   ($C_\ell$), }\,\,\,\,0.781^{+0.031}_{-0.032}\,\,(0.799) \text{ 
   ($\xi_\pm$), }\\
        \sigma_8\,:\,\,0.843^{+0.043}_{-0.020}\,\,(0.839) \text{ 
   ($C_\ell$), }\,\,\,\,0.833^{+0.042}_{-0.028}\,\,(0.851) \text{ 
   ($\xi_\pm$), }\\
   \Omega_m\,:\,\,0.277^{+0.018}_{-0.029}\,\,(0.292) \text{ 
   ($C_\ell$), }\,\,\,\,0.264^{+0.013}_{-0.028}\,\,(0.265) \text{ 
   ($\xi_\pm$). }
    \end{gathered}
\end{equation}

\begin{figure}[!ht]
\centering
\includegraphics[width=\linewidth]{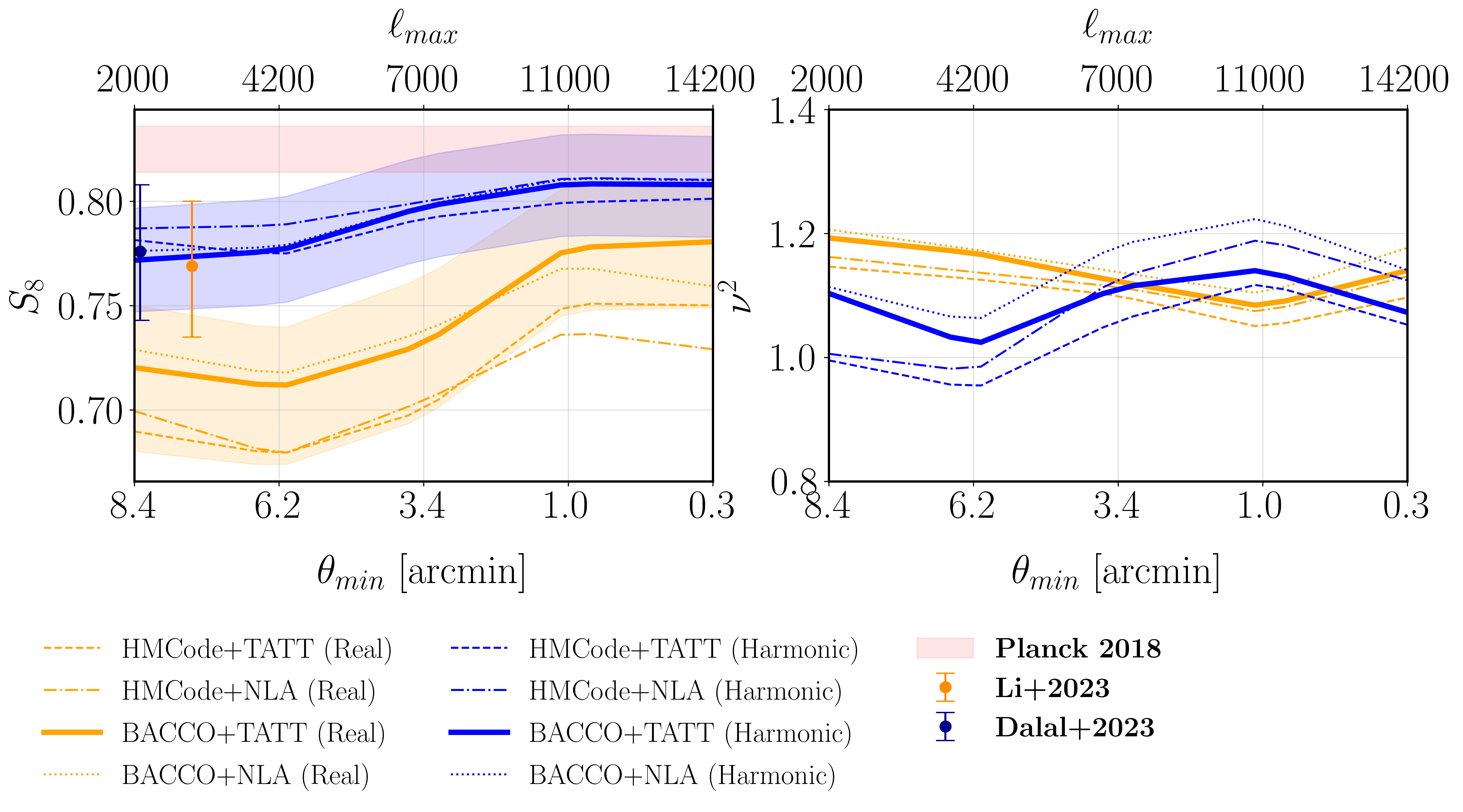}
\caption{\textbf{Left:} evolution of $S_8$ constraints for all modeling scenarios. Orange lines correspond to analyses in real space, whereas blue lines correspond to harmonic space. From left to right, we include more data points in the analysis (lower $\theta_{\rm min}$ or higher $\ell_{\rm max}$). For clarity, we show only the $68\%$ confidence interval of \textbf{BACCO+TATT}, our baseline scenario (the other scenarios have uncertainties of comparable magnitude). The red line indicates Planck 2018's best-fit value. \textbf{Right:} evolution of the reduced chi-squared for different models.
}
\label{fig:S8_evolution}

\end{figure}

From Figure~\ref{fig:whisker}, it is evident that even when including the smallest scales, the \textbf{BACCO} constraints remain overall more consistent with both official analyses. Figure~\ref{fig:S8_evolution} highlights the main feature of our results: an increase in $S_8$ as we include smaller scales, across all modeling scenarios. Interestingly, the real-space analyses exhibit a steeper rise in $S_8$, suggesting a greater sensitivity to scale cuts, possibly reflecting the way baryonic feedback affects $\xi_\pm$. We also observe that the precision of $S_8$ saturates beyond $\ell = 2000$ (or equivalently $\theta = 8.4'$), with no further tightening when including even smaller scales. A similar saturation occurs for the other cosmological parameters, although in those cases we achieve higher precision than the official HSC-Y3 analyses. Across all scale cuts, we find that our baseline model, \textbf{BACCO+TATT}, yields the most consistent results between real and harmonic spaces. In \appref, we show that this behavior extends to all cosmological parameters.

In real space, our \textbf{HMCode+TATT} results are consistent with T24~\cite{terasawa2024exploringbaryoniceffectsignature}: $S_8$ remains in mild tension with Planck at $\sim2\,\sigma$. However, this is not the case for the harmonic-space analysis, which gives $S_8=0.801^{+0.027}_{-0.028}$ -- a $\sim1.1\,\sigma$ shift relative to the real-space result.

\subsection*{Goodness-of-fit}
\noindent We now assess the goodness-of-fit for each analysis by computing the reduced chi-squared (right panel of Figure \ref{fig:S8_evolution}), defined as $\nu^2\equiv\chi^2/(m-n)$, where $m$ is the number of data points and $n$ is the number of model parameters (all models share the five common cosmological parameters). At the largest scales (most conservative scale cut, $\theta_{\min} = 8.4$ arcmin, $\ell_{\max} = 2000$), the $\nu^2$ values are lowest for the \texttt{HMCode} models, which show the most appropriate initial fit to the data. This regime is where the differences are most pronounced. The observation that real-space analysis exhibits a slightly higher $\nu^2$ compared to the harmonic analysis at these largest scales stem from the fact that our scale cuts on the real-space two-point correlation functions are not independently fine-tuned to completely disentangle small-scale effects present in the data. As we move to smaller scales (more aggressive scale cuts, $\theta_{\min} = 3.4$ arcmin, $\ell_{\max} = 7000$), the differences in $\nu^2$ between all models tend to converge and narrow significantly. All models achieve similarly acceptable fits, with the $\nu^2$ differences becoming minimal at the smallest scale shown ($\theta_{\min} \approx 0.3$ arcmin). While models with a potentially higher effective number of parameters might be theoretically penalized by the $\nu^2$ definition, the overall conclusion is that both $\texttt{HMCode-2016}$ and $\texttt{BACCO}$ models have similarly acceptable and robust performance from the point of view of the reduced chi-squared across the small-scale regime.

\subsection*{Impact of baryonic effects and intrinsic alignments}

\begin{figure}[ht!]
    \vspace{-1em}
    \includegraphics[width=\linewidth]{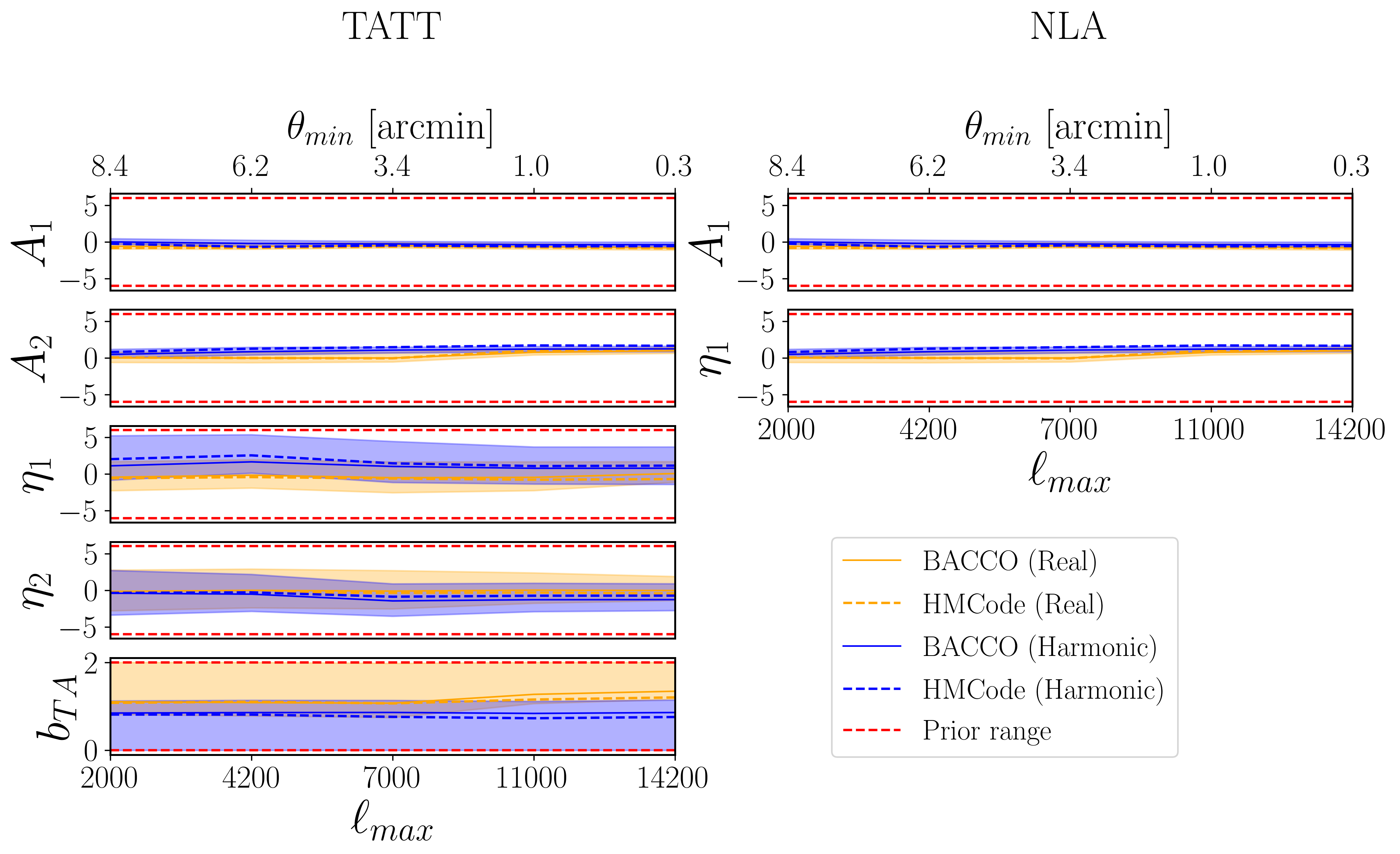}
    \caption{Evolution of TATT (left) and NLA (right) parameters for different analysis spaces and baryonic feedback models. 
    }
    \label{fig:IA_evolution}
\end{figure}

\noindent We now examine the specific impact of modeling choices for intrinsic alignment (IA) and baryonic feedback on $S_8$. To compare these choices, we quantify the systematic bias as $|S_{8,i} - S_{8,j}| / \sqrt{\sigma_i^2 + \sigma_j^2}$, representing the shift between two models relative to their combined uncertainty. Beginning with IA, Figure \ref{fig:IA_evolution} shows the evolution of NLA and TATT parameters across different spaces and baryonic feedback models. These parameters vary only mildly with scale cuts, indicating that IA lacks a strong scale-dependent signature in the current data. This is reflected in the $S_8$ estimates, where the bias introduced by the choice of IA model is approximately $0.4$--$0.5\sigma$ in real space, dropping to $0.1$--$0.2\sigma$ in harmonic space.

\begin{figure}[ht!]
    \vspace{0em}
    \includegraphics[width=\linewidth]{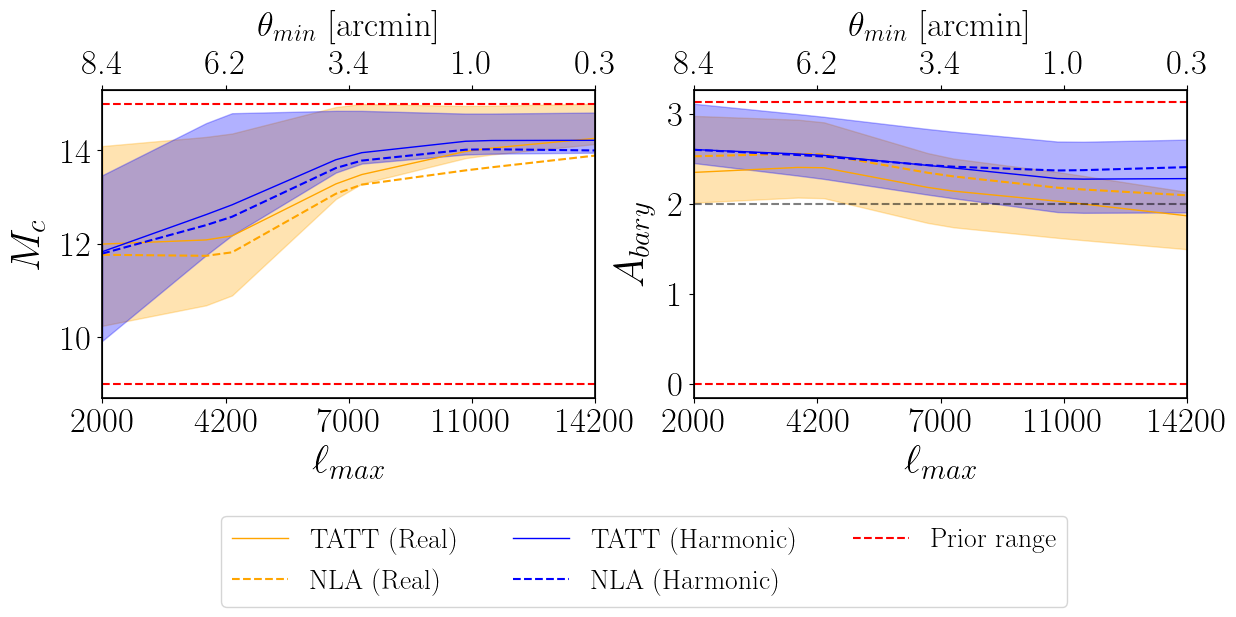}
    \caption{Evolution of $M_c$ (left) and $A_{bary}$ (right) parameters for different analysis spaces and intrinsic alignment models. In the right panel, the dashed black line represents the lower validation limit for \texttt{HMCode-2016}.
    }
    \label{fig:BF_evolution}
\end{figure}

\begin{figure}[ht!]
    \vspace{-1em}
    \includegraphics[width=1\linewidth]{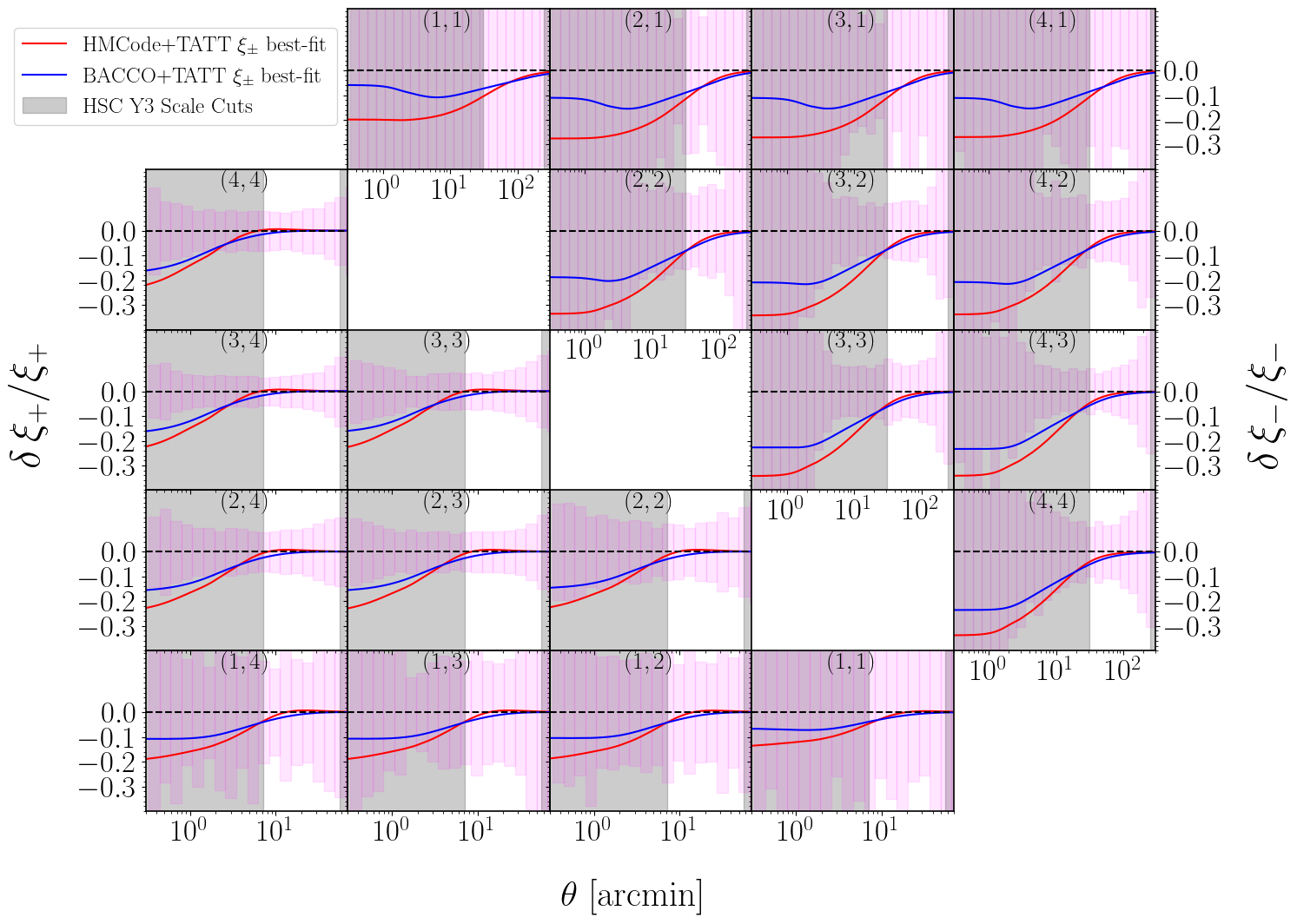}
    \caption{Suppression of DM-only $\xi_\pm$ predicted by the HMCode+TATT (red) and BACCO+TATT (blue) best fits. The magenta region marks the sensitivity range of the HSC Y3 measurements.}
    \label{fig:xi_suppression}
\end{figure}

Figure $\ref{fig:BF_evolution}$ shows the evolution of the \texttt{BACCO} $M_c$ and \texttt{HMCode-2016} $A_{\text{bary}}$ parameters across different modeling spaces and intrinsic alignment (IA) models. Our BACCO results are consistent with the Y1 results of GG24\cite{Garc_a_Garc_a_2024}: both real- and harmonic-space analyses favor strong baryonic suppression. Figure $\ref{fig:xi_suppression}$ illustrates this suppression for our best-fit $\texttt{BACCO+TATT}$ and $\texttt{HMCode+TATT}$ models in real space {($\theta_{\rm min}=0'\!.3$)}. The data prefer up to $\sim25\%$ suppression relative to the DM-only spectra (also, see the suppression at the power spectrum level in Figure \ref{fig:S_k}). This magnitude of suppression is of the same order that T24 noticed could make the HSC Y3 analysis consistent with Planck 2018 cosmological parameters. The crucial difference between the T24 analysis and this work is that we extrapolated \texttt{HMCode-2016} baryon model which has no upturn feature, while T24 used updated \texttt{HMCode-2020} baryon model which reproduce upturn at high-$k$ found in the hydrodynamical simulations.
Although such levels of suppression are supported by some hydrodynamical simulations (\cite{Le_Brun_2014,Vogelsberger_2014,stu1654}), X-ray observations and its combination with kinetic/thermal Sunyaev-Zeldovich effects (\cite{xray1,xray0,xray2,xray3,xray4,xray5,xray6,xray8,xray9,xray13,xray12,xray11,xray10}), these results should be interpreted with caution. At the same time, while a large portion of current analyses point to mild suppression, this conclusion is highly sensitive to both the priors and baryonic feedback calibrations adopted in the analysis -- as demonstrated by the distinct results in, e.g., Xu et al. \cite{xu2025constrainingbaryonicfeedbackcosmology} and Bigwood et al. \cite{bigwood2025confrontingcosmicshearastrophysical}.

For the \texttt{HMCode} case, we find that the real-space constraints on $A_{bary}$ are systematically lower across all scale cuts compared to the harmonic-space constraints. On the other hand, for the \texttt{BACCO} case, the real- and harmonic-space constraints on $M_c$ are consistent with each other among all scale cuts.
This suggests that marginalizing over a more flexible baryonic feedback model can 
improve consistency between real- and harmonic-space analyses, as also noted by GG24. We examine this hypothesis in a later subsection.

\begin{figure}[ht!]
    \hspace{2em}
    \includegraphics[width=0.8\linewidth]{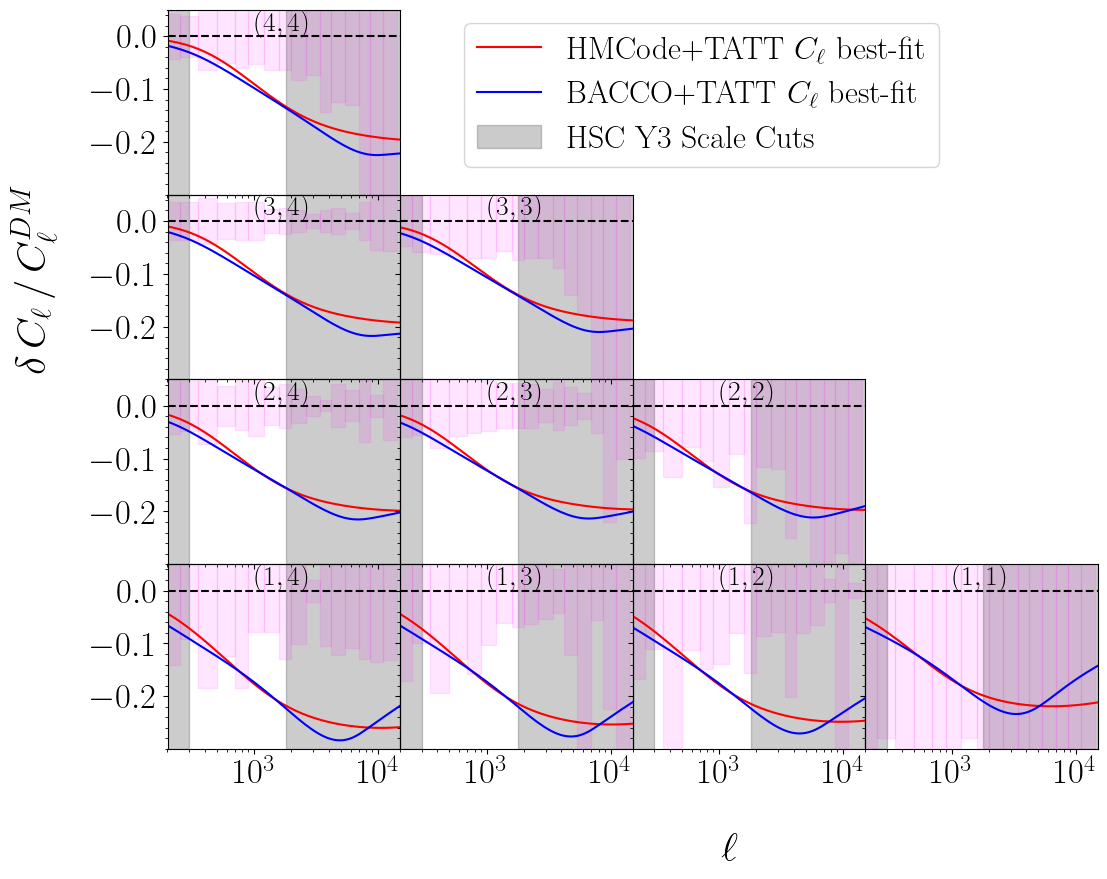}
    \caption{Suppression of DM-only $C_\ell$ predicted by the HMCode+TATT (red) and BACCO+TATT (blue) best fits. The magenta region marks the sensitivity range of the HSC Y3 measurements.}
    \label{fig:cl_suppression}
\end{figure}

Regarding the biasing on $S_8$ due to baryonic feedback modeling choices, we find a similar trend to that of the IA choices: we find a bias of $\sim 0.7\sigma$ in real space versus $\sim 0.2\sigma$ in harmonic space. These results indicate that harmonic-space analyses are generally two to three times less sensitive to small-scale modeling choices than real-space analyses -- at least given current data precision. Figure \ref{fig:cl_suppression} illustrates this pattern; as the local features introduced by IA and baryonic feedback are redistributed across a range of multipoles, model-dependent peculiarities are smoothed out. This is evident in the best-fit curves, where the suppression patterns for \texttt{BACCO} and \texttt{HMCode-2016} appear remarkably similar in harmonic space.

\subsection*{Which $\xi$ component drives the constraints}

\noindent Because $\xi_+$ is dominated by large-scale correlations with higher amplitude, it generally exhibits a higher signal-to-noise ratio than $\xi_-$, which probes smaller and noisier scales. However, as the differences we observe in the cosmological parameters appear to originate from small-scale effects, it is not immediately clear whether $\xi_-$ or $\xi_+$ is driving these discrepancies. We therefore re-analyzed the \textsc{HMCode+TATT} scenario in two complementary ways to assess which real-space component, $\xi_+$ or $\xi_-$, drives the cosmological constraints. In the first case, we retained the official HSC-Y3 scale cuts for $\xi_-$ while extending $\xi_+$ to smaller angular scales. In the second, we kept the official $\xi_+$ cuts and instead varied the $\xi_-$ cuts toward smaller scales. In both cases, including more small-scale data shifts the inferred $S_8$ value upward, but the first configuration -- where additional $\xi_+$ information is included -- shows a stronger sensitivity to the cuts. This behavior is consistent with expectations, as $\xi_+$ carries a higher signal-to-noise ratio and thus exerts a greater influence on the resulting parameter constraints.

It is also important to note that baryonic feedback affects the two correlation functions differently. In our modeling, the suppression induced by baryonic physics reaches approximately $15\%$ for $\xi_+$, but up to $25\%$ for $\xi_-$ on comparable angular scales. The fact that the current $\xi_-$ measurements appear consistent with such a strong suppression may suggest that this component remains relatively forgiving: its lower signal-to-noise reduces its constraining power, allowing it to accommodate a wider range of baryonic scenarios. As pointed out by T24, however, the higher $\xi_-$ signal-to-noise expected in upcoming surveys will be crucial to better constrain baryonic suppression models.

\subsection*{On the theoretical $\xi_{\pm}$}
\noindent As discussed in Section \ref{subsec:scale_cuts}, a proper characterization of $\xi_\pm(\theta_{\rm min})$ requires integrating Eqs.~\eqref{eq:xi_plus}-\eqref{eq:xi_minus} up to a multipole $\ell_{\rm max}\!\sim\!10\,\pi\,/\,\theta_{\rm min}$ (for a $\sim1\%$ error estimate). However, as our signal-to-noise ratio drops at smaller scales, the signal from such high multipoles becomes increasingly suppressed. We tested this hypothesis and found a bias of $0.1$-$0.2\,\sigma$ in our constraints when integrating up to $\ell_{\rm max}=\,\pi\,/\,\theta_{\rm min}$ instead of $\ell_{\rm max}=10\,\pi\,/\,\theta_{\rm min}$. This difference lies well within the numerical uncertainty introduced by integrating $\xi_\pm$ over logarithmically spaced $\ell$ intervals.

\subsection*{On the impact of the sampler}

\begin{figure}[ht!]
    \hspace{1em}
    \includegraphics[width=0.9\linewidth]{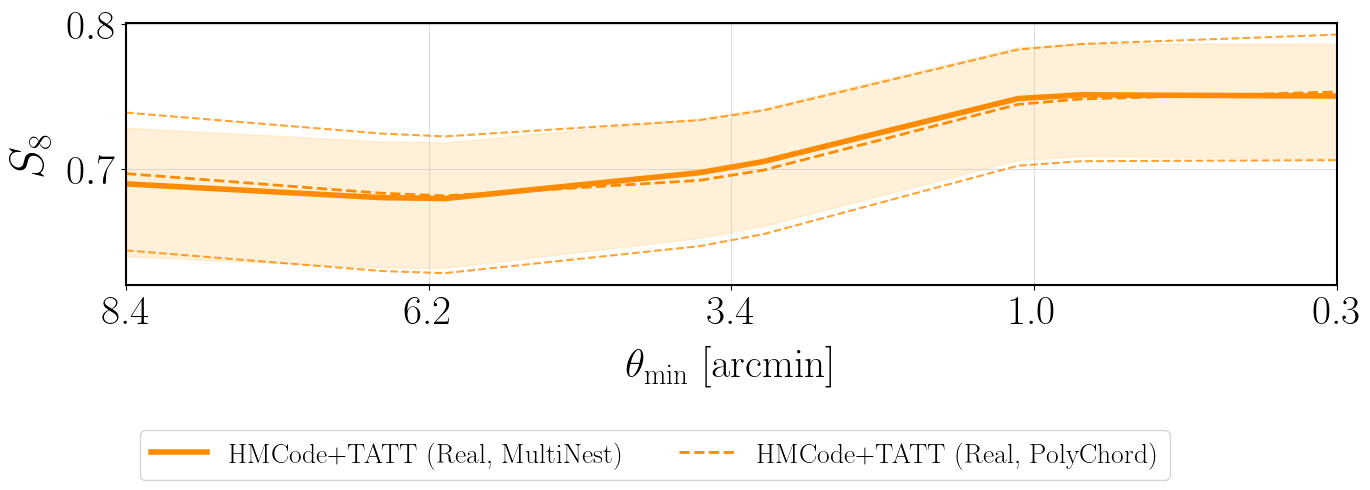}
    \caption{Comparison between \texttt{MultiNest} (orange) and \texttt{PolyChord} (blue) nested samplers. Apart from a slight underestimation of parameter uncertainties, both samplers yield consistent constraints.}
    \label{fig:mn_pc}
\end{figure}

\noindent All of our previous results were obtained using the \texttt{CosmoSIS} nested sampler, which typically underestimates parameter uncertainties for the HSC-Y3 data compared to other nested samplers and standard MCMC methods. To also investigate potential biases in our inferred parameters, we re-performed the \textbf{HMCode+TATT} analysis in real space using the \texttt{PolyChord} nested sampler, which is roughly four times slower than \texttt{MultiNest} but yields constraints that are consistent with those from MCMC approaches. Figure~\ref{fig:mn_pc} shows that, apart from the underestimation of parameter errors, there is no evident bias in our $S_8$ results.

\subsection*{On the marginalization over \texttt{BACCO} parameters}

\noindent We explored whether the observed improvements in consistency and the upward shift in $S_8$ were attributable to the full 7-parameter flexibility of the $\texttt{BACCO}$ model, or primarily to its central parameter, $M_c$. The parameter $M_c$ sets the halo mass scale above which baryonic feedback significantly alters density profiles. To test this, we re-ran the \textbf{BACCO+TATT} analyses (both real-space and harmonic-space) while marginalizing only over $M_c$, keeping the other six $\texttt{BACCO}$ parameters fixed at their best-fit values from the original 7-parameter analysis. We found that the resulting constraints on $S_8$ and all other cosmological parameters were highly consistent with those from the full 7-parameter marginalization. This strongly indicates that the single parameter $M_c$ provides sufficient flexibility to capture the essential scale-dependent suppression of the matter power spectrum due to baryonic effects that is constrained by the HSC Y3 two-point statistics. The remaining six parameters, which control finer details of the baryonic redistribution, are weakly constrained given the current data precision. Therefore, the improved robustness and higher $S_8$ are a robust consequence of adopting a more flexible suppression scale, primarily governed by $M_c$.

\section{Conclusions}
\label{sec:conclusions}

\noindent In this work, we investigated the consistency of cosmic shear analyses performed in real space ($\xi_\pm$) and harmonic space ($C_\ell$) using HSC Y3 data, focusing on the inclusion of small-scale information. We progressively extended our analysis to smaller angular scales ($\theta_{\rm min} = 0.3'$) and higher multipoles ($\ell_{\rm max} = 14200$), testing the impact of different models for Intrinsic Alignment (NLA, TATT) and baryonic feedback (\texttt{HMCode-2016}, \texttt{BACCO}).

Our results confirm that real and harmonic spaces are intrinsically different probes regarding their sensitivity to non-linear physics. As shown in Figures \ref{fig:xi_suppression} and \ref{fig:cl_suppression}, a sharp feature in one space -- such as the suppression of power by baryonic feedback -- is "washed out" and smeared across a wide range of scales in the other due to the nature of the Bessel transform. This makes the harmonic-space analysis inherently less sensitive to the specific modeling of small-scale physics; we found that the bias in $S_8$ from switching between baryonic or IA models was two to three times smaller in the $C_\ell$ analysis compared to the $\xi_\pm$ analysis.

This finding highlights a key argument of our paper: instead of fine-tuning scale cuts to force consistency, we should orient ourselves toward building a theoretical model that is robust and reliable in both spaces simultaneously. We found that the standard \texttt{HMCode+TATT} model, while fitting the data, yields a significant $\sim 1.1\sigma$ tension in $S_8$ between the full real-space and harmonic-space analyses. In contrast, the more flexible \texttt{BACCO+TATT} model provides the more consistent constraints between the two spaces across all scale cuts (Fig. \ref{fig:S8_evolution}). We further demonstrated that this robustness does not require the full complexity of the emulator's parameter space; marginalizing solely over the characteristic mass parameter $M_c$ is sufficient to reproduce the results where all the parameters are marginalized.

We find a clear complementarity between the two spaces. For current surveys, the harmonic-space power spectrum, being less sensitive to the sharp features of baryonic feedback, appears to be a more robust probe for constraining cosmological parameters. However, this same insensitivity makes it difficult to distinguish between different baryonic feedback models. As we move to Stage-IV surveys, the high signal-to-noise measurements in real space--particularly for $\xi_{-}$--will be essential for precisely characterizing baryonic physics. Therefore, real-space analyses will be indispensable for validating the astrophysical models required to unlock the full cosmological potential of future surveys.

Finally, while we extended our analyses to the smallest scales available, we stress that these results should be interpreted with care. Figure~\ref{fig:S8_evolution} shows a systematic shift in $S_8$ as progressively smaller scales are included, suggesting that the constraints may be biased if the baryonic modeling is incomplete. Nonetheless, this shift is substantially less pronounced in harmonic space, largely independent of the modeling choice. This robustness indicates that harmonic-space analyses may offer a safer path for exploiting the high signal-to-noise at small scales, even when baryonic physics is not perfectly modeled. Future surveys may therefore benefit from relying more heavily on $C_\ell$ at the smallest scales, using real-space statistics to calibrate baryonic feedback models.

\section*{Acknowledgments}

\noindent The authors would like to thank Masahiro Takada, Joaquin Armijo and Raul Abramo for their insights and valuable comments. JF is supported by the Coordenação de Aperfeiçoamento de Pessoal de Nível Superior (CAPES) Grant 88881.982198 and the National Council for Scientific and Technological Development (CNPq) Grant 132397. RT is supported by JSPS KAKENHI Grant 23KJ0747.

\bibliographystyle{JHEP}
\bibliography{stuff.bib}

\clearpage
\appendix
\section*{Appendix: Complementary Tables and Figures}
\label{app:A}
\begin{table}[ht!]
\label{tab:priors}
\centering
\caption{Model parameters and priors used in the fiducial cosmological inference following \cite{li2023hypersuprimecamyear3}. “U($a,b$)” denotes a uniform prior between $a$ and $b$, and “N($\mu,\sigma$)” denotes a normal prior with mean $\mu$ and standard deviation $\sigma$.
}
\vspace{1em}
\resizebox{0.6\textwidth}{!}{%
\begin{tabular}{l l l}
\hline\hline
\textbf{Cosmological parameters} & \\
$\Omega_m$ & U(0.1,0.7) \\
$A_s\,\,( \times10^{-9})$ & U(0.5,10)  \\
$n_s$ & U(0.87,1.07)  \\
$h_0$ & U(0.62,0.80)  \\
$\omega_b$ & U(0.02,0.025)  \\[2pt]
\textbf{Baryonic feedback (\texttt{HMCode-2016})}  &  \\
$A_b$ & U(2,3.13) &  \\[2pt]
\textbf{Baryonic feedback (\texttt{BACCO})}  &  \\
$\rm{log}_{10}\,M_c$ & U(9,15) \\[2pt]
$\rm{log}_{10}\,\eta$ & U(-0.698,0.698) \\[2pt]
$\rm{log}_{10}\,\beta$ & U(-1.000,0.698) \\[2pt]
$\rm{log}_{10}\,M_{1,z_0,cen}$ & U(9,13) \\[2pt]
$\rm{log}_{10}\,\theta_{out}$ & U(0.000, 0.477) \\[2pt]
$\rm{log}_{10}\,\theta_{inn}$ & U(-2.000,-0.522) \\[2pt]
$\rm{log}_{10}\,M_{inn}$ & U(9.0,13.5) \\[2pt]
\textbf{Intrinsic alignment (TATT)} & \\
$A_1$ & U($-6$,6) \\
$\eta_1$ & U($-6$,6) \\
$A_2$ & U($-6$,6) \\
$\eta_2$ & U($-6$,6) \\
$b_{ta}$ & U(0,2) \\[2pt]
\textbf{Intrinsic alignment (NLA)} & \\
$A_1$ & U($-5$,5) \\
$\eta_1$ & U($-6$,6) \\
\textbf{Photo-$z$ systematics} & \\
$\Delta z_1$ & N(0,0.024) \\
$\Delta z_2$ & N(0,0.022) \\
$\Delta z_3$ & U($-1$,1) \\
$\Delta z_4$ & U($-1$,1) \\[2pt]
\textbf{Shear calibration biases} & \\
$\Delta m_1$ & N(0.0,0.01) \\
$\Delta m_2$ & N(0.0,0.01) \\
$\Delta m_3$ & N(0.0,0.01) \\
$\Delta m_4$ & N(0.0,0.01) \\[2pt]
\textbf{PSF systematics} & \\
$\alpha'^{(2)}$ & N(0,1) \\
$\beta'^{(2)}$ & N(0,1) \\
$\alpha'^{(4)}$ & N(0,1) \\
$\beta'^{(4)}$ & N(0,1) \\[2pt]
\hline
\end{tabular}
}
\end{table}

\clearpage
\begin{figure}[ht!]
    \vspace{-1em}
    \includegraphics[width=1\linewidth]{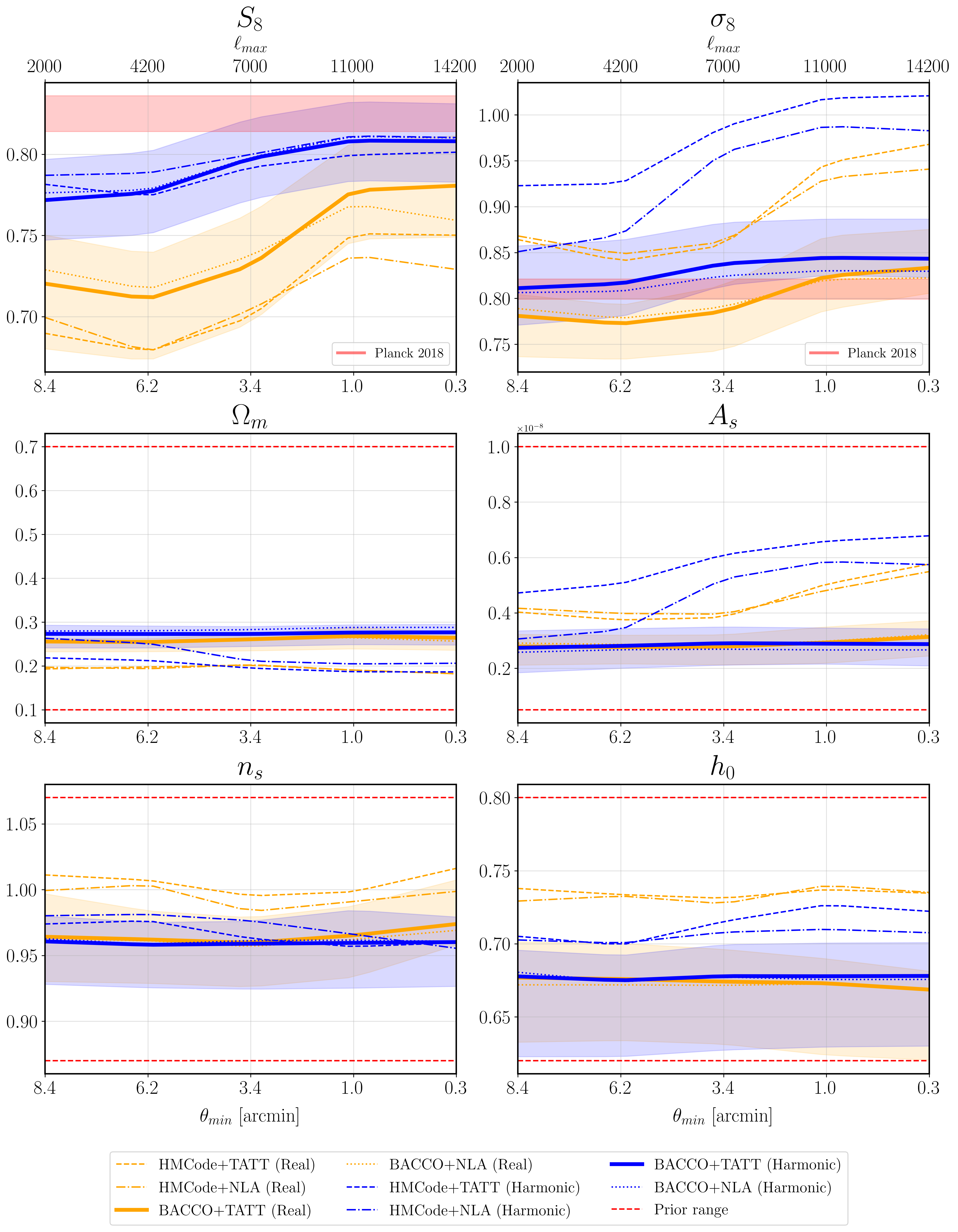}
    \caption{Evolution of cosmological constraints for all modeling scenarios. Orange lines corresponds to analyses in real space, whereas blue corresponds to harmonic space. From left to right, we include more data points into the analysis (lower $\theta_{\rm min}$ or higher $\ell_{\rm max}$). For better visualization, we only plot the $68\%$ confidence interval of \textbf{BACCO + TATT}, our baseline scenario (other scenarios have uncertainties of the same order of magnitude). We also show Planck 2018's best-fit in red.}
    \label{fig:sigma8_evolution}
\end{figure}

\clearpage
\label{app:fiducial_triangle}
\begin{figure}[ht!]
    \centering
    \includegraphics[width=\textwidth]{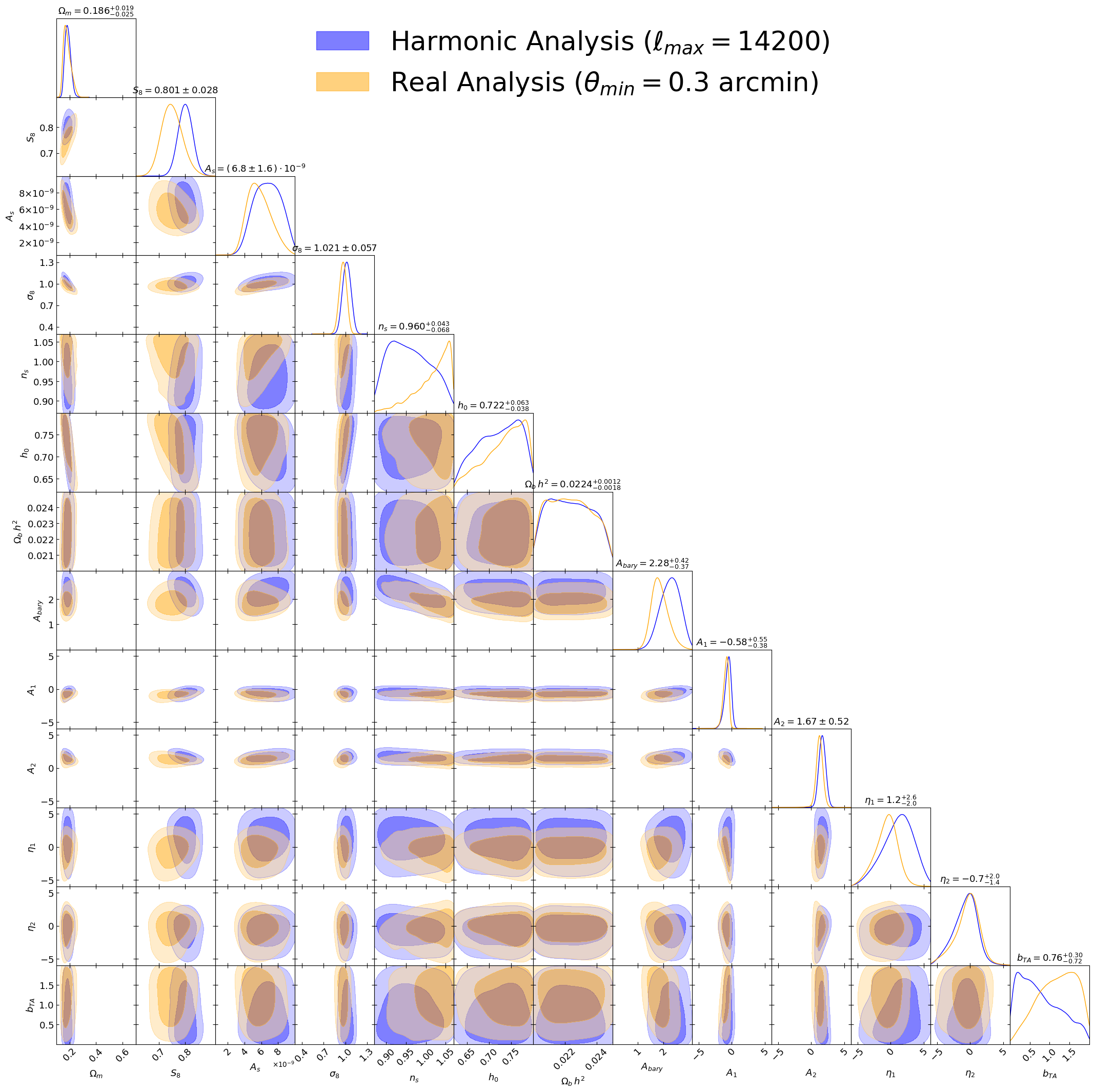}
    \caption{Constraints in cosmological, baryonic feedback and intrinsic alignment parameters ($1\sigma$ and $2\sigma$ contours), for the \textbf{(HMCode+TATT) }scenario. The orange contours correspond to the real-space analysis including all available data from HSC Y3 ($\theta_{\mathrm{min}}=0.3'$). The blue contours correspond to the analogous analysis in harmonic space ($\ell_{\mathrm{max}}=14200$). For reference, the best-fit values for the real-space analysis are shown on top of each 1D posterior.}

\end{figure}

\clearpage

\begin{figure}[ht!]
    \centering
    \includegraphics[width=\textwidth]{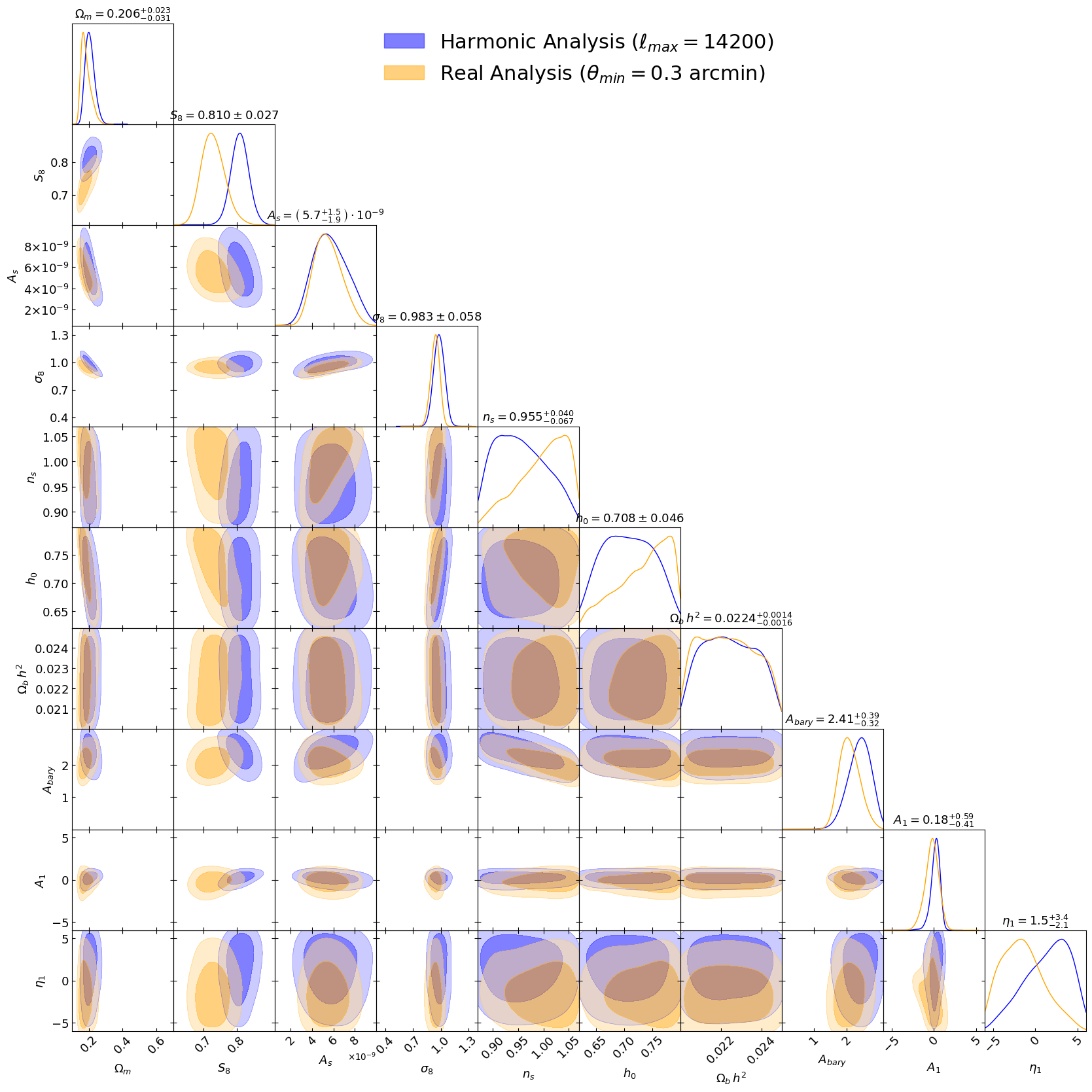}
    \caption{Constraints in cosmological, baryonic feedback and intrinsic alignment parameters ($1\sigma$ and $2\sigma$ contours), for the \textbf{(HMCode+NLA) }scenario. The orange contours correspond to the real-space analysis, including all available data from HSC Y3 ($\theta_{\mathrm{min}}=0.3'$). The blue contours correspond to the analogous analysis in harmonic space ($\ell_{\mathrm{max}}=14200$). For reference, the best-fit values for the real-space analysis are shown on top of each 1D posterior.}

\end{figure}

\clearpage

\begin{figure}[ht!]
    \centering
    \includegraphics[width=\textwidth]{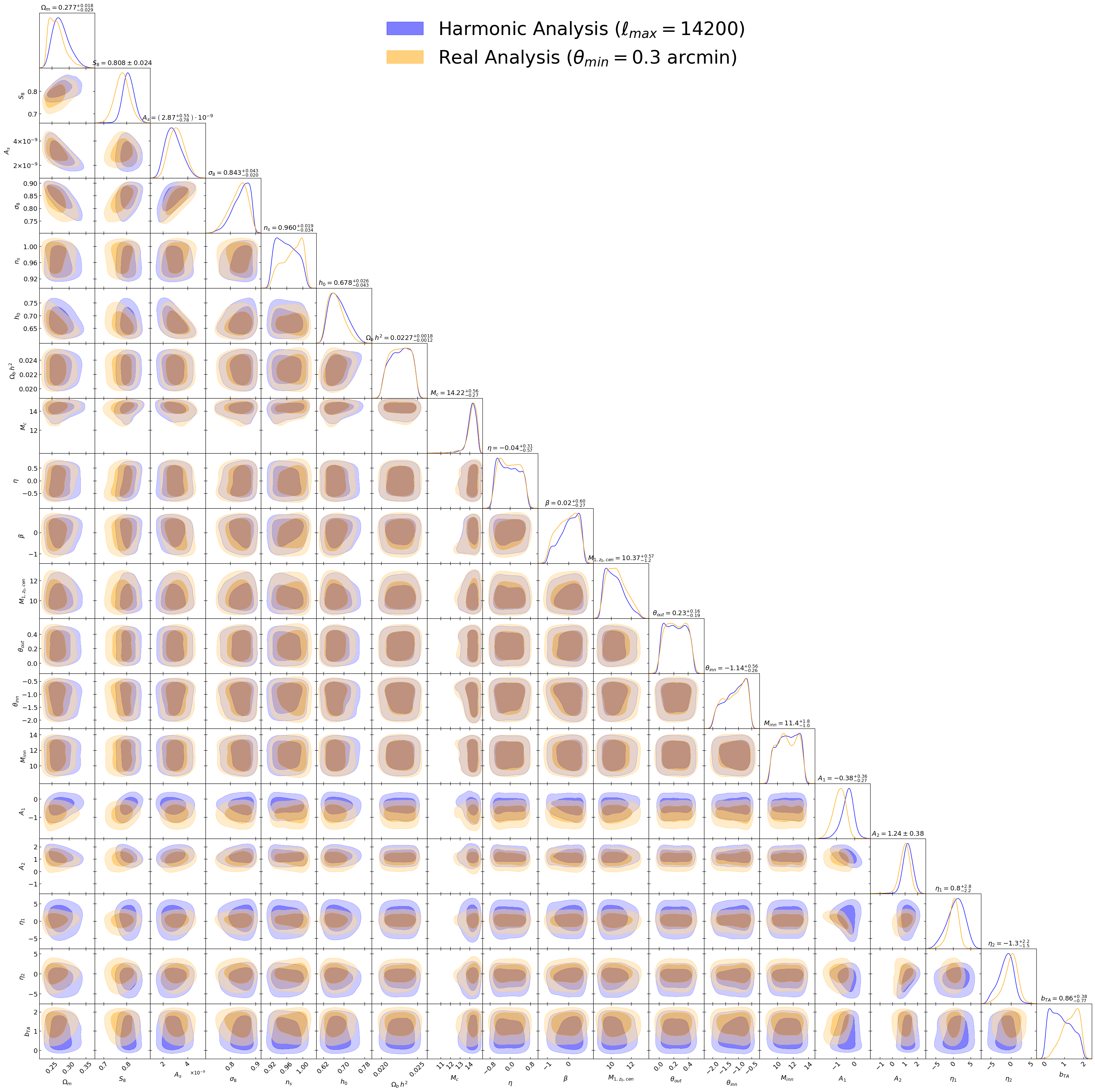}
    \caption{Constraints in cosmological, baryonic feedback and intrinsic alignment parameters ($1\sigma$ and $2\sigma$ contours), for the \textbf{(BACCO+TATT) }scenario. The orange contours correspond to the real-space analysis including all available data from HSC Y3 ($\theta_{\mathrm{min}}=0.3'$). The blue contours correspond to the analogous analysis in harmonic space ($\ell_{\mathrm{max}}=14200$). For reference, the best-fit values for the real-space analysis are shown on top of each 1D posterior.}

\end{figure}

\begin{figure}[ht!]
    \centering
    \includegraphics[width=\textwidth]{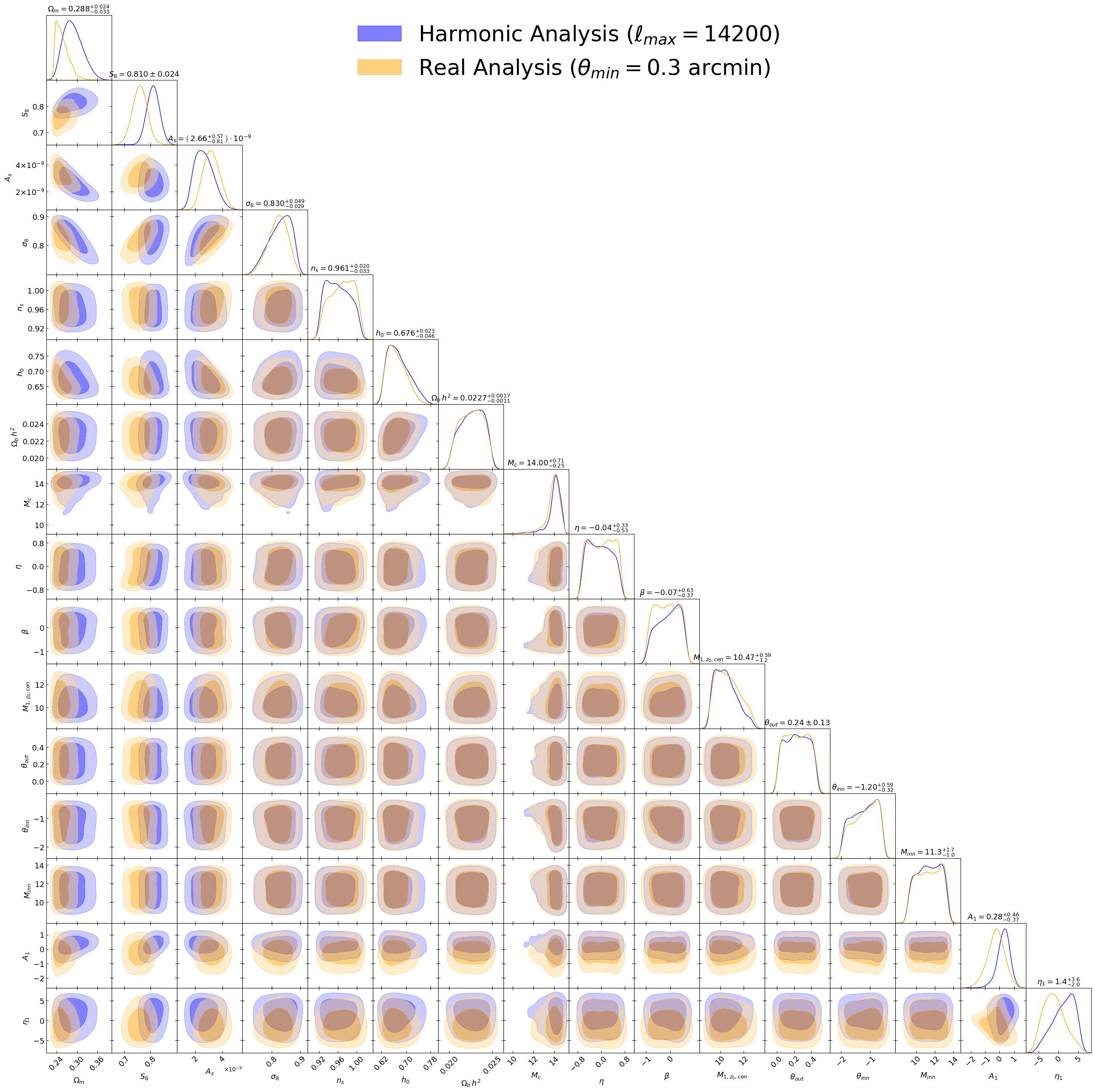}
    \caption{Constraints in cosmological, baryonic feedback and intrinsic alignment parameters ($1\sigma$ and $2\sigma$ contours), for the \textbf{(BACCO+NLA) }scenario. The orange contours correspond to the real-space analysis including all available data from HSC Y3 ($\theta_{\mathrm{min}}=0.3'$). The blue contours correspond to the analogous analysis in harmonic space ($\ell_{\mathrm{max}}=14200$). For reference, the best-fit values for the real-space analysis are shown on top of each 1D posterior.}
    \label{fig:fiducial_triangle}
\end{figure}

\begin{figure}[ht!]
    \centering
    \includegraphics[width=1\linewidth]{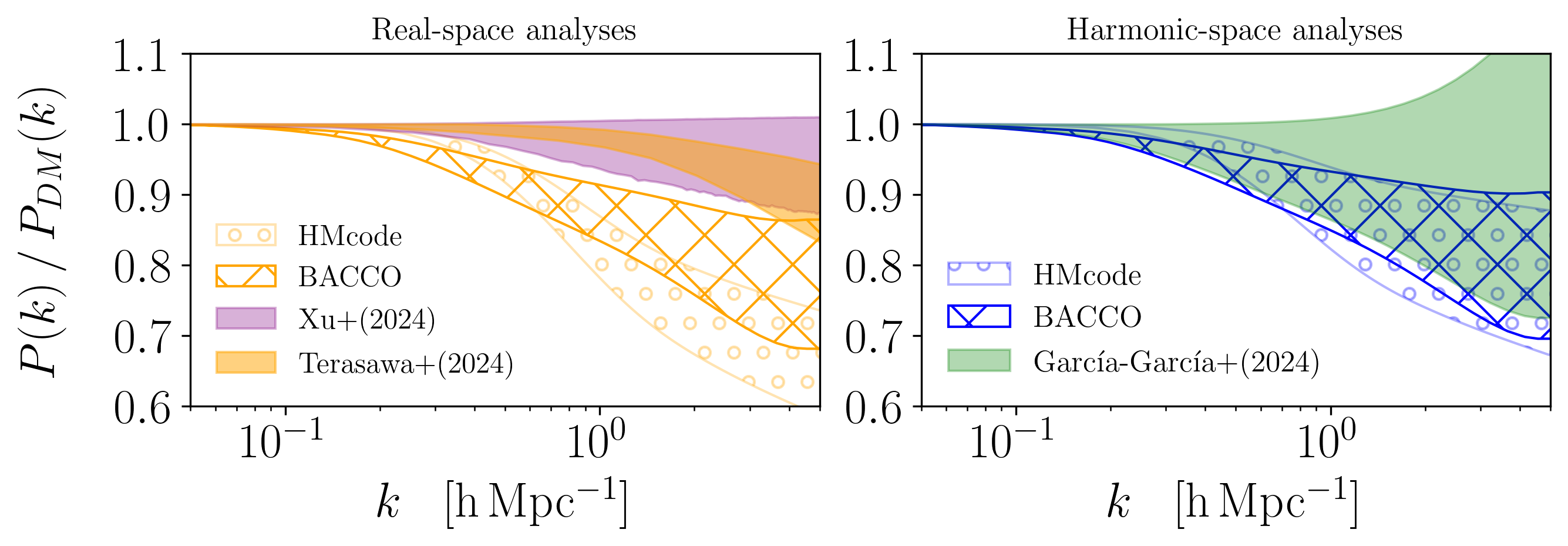}
    \caption{Baryonic suppression at the power spectrum level. Left: suppressions inferred from real-space analyses, in particular this paper's \textbf{HMCode+TATT} and \textbf{BACCO+TATT} results {($\theta_{\rm min}=0'\!.3$)}. For comparison, we also show recent suppression constraints from the literature: the Xu et al.\ \cite{xu2025constrainingbaryonicfeedbackcosmology} 2PCF results (purple) from the DES Y3 $\times$ Planck DR4 6$\times$2pt analysis ($\theta_{\min}=2.5^\prime$), and the T24 \cite{terasawa2024exploringbaryoniceffectsignature} 2PCF results for HSC-Y3 ($\theta_{\min}=0.3^\prime$) shown in shaded orange. Right: suppressions inferred from harmonic-space analyses, highlighting this paper's \textbf{HMCode+TATT} and \textbf{BACCO+TATT} results 
    {($\ell_{\rm max}=14200$)}
    alongside the recent constraints from García-García et al.\cite{Garc_a_Garc_a_2024} (green), obtained by combining HSC DR1, DES Y3, and KiDS-1000 up to $\ell_{max}=8192$.
    } 
    \label{fig:S_k}
\end{figure}

\end{document}